\documentclass[11pt]{article}
\usepackage[utf8]{inputenc}
\usepackage{graphicx}
\usepackage{overpic}
\usepackage{amsmath, amssymb, amsthm}
\usepackage{geometry}
\usepackage{booktabs}
\usepackage{xcolor}
\usepackage[numbers,sort&compress]{natbib}
\usepackage[colorlinks=true,linkcolor=blue!60!black,citecolor=green!40!black,urlcolor=blue!60!black]{hyperref}
\usepackage{cleveref}
\usepackage{placeins}

\graphicspath{{figures/}}

\usepackage{tikz}
\usetikzlibrary{arrows.meta,matrix}

\usepackage{xcolor}

\definecolor{acc}{HTML}{2878B5}
\definecolor{cR}{HTML}{D55E00}

\usepackage{pgfplots}
\usepgfplotslibrary{groupplots}
\pgfplotsset{compat=1.18}

\title{Parametric spatio-temporal reduced-order modelling via \\
Spectral Proper Orthogonal Decomposition and Gaussian Process Regression}
\author{Guglielmo Padula\textsuperscript{1,}\thanks{gpadula@sissa.it} \quad Antonio Colanera\textsuperscript{1,}\thanks{Corresponding author. acolaner@sissa.it} \quad Gianluigi Rozza\textsuperscript{1,}\thanks{grozza@sissa.it}\\[4pt]
\small \textsuperscript{1}SISSA MathLab, Mathematics Area, SISSA, via Bonomea 265, I-34136 Trieste, Italy}
\date{}

\begin{document}

\maketitle

\begin{abstract}
We present SPOD-GPR, a non-intrusive, parametric, spatio-temporal reduced-order model (ROM) that combines Proper Orthogonal Decomposition (POD), Spectral POD (SPOD), and Gaussian Process Regression (GPR). A parametric POD first projects the high-dimensional state onto a low-dimensional subspace. The reduced trajectories are then divided into windowed sections, transformed into the frequency domain, and truncated to the most energetic frequencies. SPOD is then performed independently at each retained frequency, yielding temporally coherent modes and coefficients. The parametric dependence of the POD and SPOD bases is modelled by GPR on a local chart of the Grassmann manifold after Procrustes alignment. The SPOD modal coefficients are learned both as explicit functions of the parameters and through a parameter-conditioned autoregressive model that allows for temporal extrapolation beyond the training window. Gaussian processes are sampled jointly and pathwise via Matheron conditioning, ensuring that each predictive realization carries uncertainty quantification consistent with the posterior distribution.
The method is tested on four parametric examples where the parameter values and the prediction horizon lie outside the training data: the two-dimensional anisotropic Kuramoto--Sivashinsky equation in a time-periodic and a travelling-wave regime, each parametrised by one of its two coefficients; flow past a cylinder parametrised by the Reynolds number; and flow past an elliptic obstacle parametrised by its semi-axes. In all cases, SPOD-GPR reduces the normalised mean absolute error compared to a POD-GPR baseline, and improves the calibration of the predictive uncertainty, especially in the fluid flow cases, where POD-GPR is severely overconfident. This opens the way to spectral, uncertainty-aware reduced-order models for many-query and real-time applications, such as digital twins.
\end{abstract}


\section{Introduction}
\label{sec:intro}

Applications that involve a large number of queries or require real-time performance, such as design optimisation, control, uncertainty quantification, and the use of digital twins, rely on solving parametrised, time-dependent partial differential equations (PDEs) repeatedly. Reduced-order models (ROMs) meet this demand by replacing the high-fidelity full-order model (FOM) with a low-dimensional surrogate~\cite{hesthaven_certified_2016, padula_brief_2024}. Unlike projection-based (intrusive) ROMs, which require access to the discretised operators and are often impractical when the FOM is provided as legacy software or in the form of a commercial package, nonintrusive and data-driven ROMs determine the reduced dynamics directly from snapshot data~\cite{guo_reduced_2018, guo_data-driven_2019, fresca_pod-dl-rom_2022}. A typical procedure involves first reducing the snapshots using Proper Orthogonal Decomposition (POD)~\cite{lumley_stochastic_1970, berkooz_proper_1993}, and then fitting the POD coefficients as functions of the parameters and/or time; in this case, Gaussian Process Regression (GPR)~\cite{rasmussen_gaussian_2005} is especially attractive since it is data-efficient, offers strong approximation guarantees in the relevant reproducing kernel Hilbert space~\cite{wendland_scattered_2004, kanagawa_gaussian_2018, micchelli_universal_2006, maddalena_deterministic_2021}, and provides a predictive variance without having to add any extra modelling effort~\cite{guo_reduced_2018, guo_data-driven_2019}.

Gaussian processes have been widely used in the field of learning dynamical systems from data, whether for modelling one-step transition functions~\cite{wang_gaussian_2008, deisenroth_gaussian_2015}, as a way of representing vector fields for ordinary differential equations whose forms are unknown~\cite{heinonen_learning_2018, ensinger_exact_2024, padula_quadratic_2026}, or as priors that are designed to take into account the governing equations~\cite{raissi_machine_2017, raissi_numerical_2018, chen_solving_2021}. The kernel-based methods for learning dynamics have also been investigated from the point of view of cross-validation and kernel flows~\cite{owhadi_kernel_2019, hamzi_learning_2021}. Other linear and sparse-regression methods include Dynamic Mode Decomposition (DMD) and various related variants~\cite{williams_datadriven_2015, sashidhar_bagging_2022, baddoo_kernel_2022, andreuzzi_dynamic_2023}, as well as sparse identification of nonlinear dynamical systems~\cite{brunton_discovering_2016}.

A major disadvantage of most POD-based nonintrusive ROMs is that the POD modes are arranged purely according to their energy and do not show how the dynamics are organised in time. In flow situations characterised by travelling waves, quasi-periodic motions, or the presence of several coherent time scales, one POD mode can combine features associated with different frequencies; consequently, its modal coefficient shows a complex, multiscale temporal behaviour which is harder to learn and, in particular, more difficult to extrapolate. Spectral Proper Orthogonal Decomposition (SPOD), which was first introduced by Lumley~\cite{lumley_stochastic_1970,moffatt_stochastic_1975} and later considered again by Towne, Schmidt and Colonius~\cite{towne_spectral_2018, schmidt_guide_2020}, overcomes this problem by building an optimal, energy-ranked basis \emph{for each frequency}. The SPOD modes are coherent both in space and in time, correspond to optimally averaged DMD modes for statistically stationary flows, and agree with resolvent modes under certain assumptions~\cite{towne_spectral_2018}. SPOD is now widely employed to study jets~\cite{schmidt_spectral_2018}, wall-bounded turbulence~\cite{abreu_spectral_2020} and multiphase flows~\cite{colanera_modal_2021}, and has given rise to a wide range of algorithms, including streaming~\cite{schmidt_efficient_2019}, multitaper~\cite{schmidt_spectral_2022}, gappy~\cite{nekkanti_gappy_2023} and robust~\cite{colanera_robust_2025} versions, as well as methods for low-rank reconstruction and denoising~\cite{nekkanti_frequencytime_2021}. It should be noted that a separate filter-based method with the same name was put forward in~\cite{sieber_spectral_2016}. In the context of reduced-order modelling, the SPOD coefficients have been predicted by means of neural networks~\cite{lario_neural-network_2022}, showing that the frequency-resolved latent representation can be easier to learn than the POD version.

In this paper we develop these concepts and present a \emph{parametric}, \emph{probabilistic} spatio-temporal ROM which combines SPOD with GPR. The main contributions are:
\begin{enumerate}
\item We justify our use of SPOD for the purpose of learning dynamics by means of a time-delay (block) augmentation of the state: even though such an augmentation is capable of converting a non-autonomous scalar signal into an autonomous system, when POD is applied to the augmented snapshot matrix (this is known as block POD) it entangles the spatial and temporal scales; carrying out a windowed Fourier transform along the delay coordinate separates these scales and yields SPOD (\Cref{sec:motivation}).
\item We present a two-stage reduction procedure, first POD and then SPOD, enabling the application of large SPOD windows to high-dimensional datasets, together with a frequency truncation that applies to all the training parameters (\Cref{sec:stage1,sec:stage2}).
\item We make all the bases parametric by solving separate (S)POD problems at each value of the parameter, aligning the resulting bases using orthogonal/unitary Procrustes rotations, and then performing the regression on a chart of the Grassmann manifold~\cite{edelman_geometry_1998}, centred at the Karcher mean of the training subspaces; in this way the interpolated modes remain exactly orthonormal (\Cref{sec:grassmann}).
\item We use three ridge-GP surrogates to model the SPOD modes and coefficients: one for mapping the parameters to the modes, another for mapping the parameters to the trajectories, and a third which is a parameter-conditioned autoregressive model for time extrapolation. By applying joint pathwise (Matheron) sampling~\cite{wilson_efficiently_2020} across all of the GPs, uncertainty is propagated to the physical field, and the variance scale of each GP is adjusted using the leave-one-out residuals (\Cref{sec:gpr}).
\item We assess the method on four parametric benchmark problems, looking at test parameters and prediction horizons which were not encountered during training, and we compare it with a POD-GPR baseline that employs the same regression structure (\Cref{sec:results}).
\end{enumerate}

The rest of the paper is organised as follows: \Cref{sec:setting} describes the problem setting, \Cref{sec:method} sets out the SPOD-GPR methodology, \Cref{sec:results} gives the details of the numerical experiments, \Cref{sec:discussion} looks at the limitations, and \Cref{sec:conclusions} summarises the conclusions.

\section{Problem setting}
\label{sec:setting}

Let $\Omega\subset\mathbb{R}^d$ represent a spatial domain and $\mathcal{P}\subset\mathbb{R}^{d_\mu}$ a parameter set. We consider a time-dependent parametric PDE with solution $u(t,x;\boldsymbol{\mu})$, which is spatially discretised by means of a FOM with $N_h$ degrees of freedom and evaluated at uniformly spaced time instants $t_n=t_0+n\,\delta t$. The discrete state at time $t_n$ is denoted by $\mathbf{u}_n(\boldsymbol{\mu})\in\mathbb{R}^{N_h}$, and we introduce a symmetric positive-definite weighting matrix $\mathbf{W}\in\mathbb{R}^{N_h\times N_h}$, {such as a finite-element mass matrix}, defining the discrete $L^2(\Omega)$ inner product $\langle\mathbf{u},\mathbf{v}\rangle_{\mathbf{W}}=\mathbf{u}^\top\mathbf{W}\mathbf{v}$.

The training set comprises $N_t$ snapshots for each of $N_\mu$ training parameter instances belonging to the set $\mathcal{P}_{\mathrm{train}}=\{\boldsymbol{\mu}_i\}_{i=1}^{N_\mu}\subset\mathcal{P}$,
\begin{equation}
\mathbf{U}_i=
[\mathbf{u}_0(\boldsymbol{\mu}_i),\ldots,
\mathbf{u}_{N_t-1}(\boldsymbol{\mu}_i)]
\in\mathbb{R}^{N_h\times N_t},
\qquad i=1,\ldots,N_\mu.
\end{equation}

The aim is to predict the system's state for a new parameter value $\boldsymbol{\mu}^\ast\in\mathcal{P}\setminus\mathcal{P}_{\mathrm{train}}$ over a future time horizon $\{\mathbf{u}_n(\boldsymbol{\mu}^\ast)\}_{n=N_t}^{N_t+N_{\mathrm{test}}-1}$ extending beyond the training time interval, together with an estimate of the associated predictive uncertainty. This involves carrying out parametric interpolation and temporal extrapolation at the same time, a problem that is considerably more difficult than tackling either of these tasks separately.

\section{Methodology}
\label{sec:method}

\subsection{Motivation: from time-delay augmentation to spectral modes}
\label{sec:motivation}

It is simplest to deduce a dynamical law from data when the state is \emph{autonomous}, that is, when its time derivative (or its one-step increment) depends only on the state at that moment. In reality, reduced coordinates are almost never autonomous since projecting onto a small number of modes eliminates the information needed to close the evolution equations. A common solution, one that is based on delay-embedding theorems~\cite{rand_detecting_1981}, is to extend the state by adding to it time-lagged copies of itself,
\begin{equation}
\mathbf{D}(t,\boldsymbol{\mu})=\big[\,\mathbf{u}(t,\boldsymbol{\mu})^\top,\ \mathbf{u}(t+\delta t,\boldsymbol{\mu})^\top,\ \ldots,\ \mathbf{u}(t+(N_f-1)\,\delta t,\boldsymbol{\mu})^\top\,\big]^\top\in\mathbb{R}^{N_hN_f}.
\label{eq:delay}
\end{equation}
This kind of augmentation is useful in situations where a system is not originally autoregressive but does become autoregressive after time-augmented variables have been incorporated. A simple example is the case of $y(t)=\sin t$: since $\dot y=\cos t=\pm\sqrt{1-y^2}$, the time derivative cannot be given as a function of $y$ alone and therefore no first-order autonomous scalar equation can generate this signal. On the other hand, if the state is augmented with a shift of $\pi/2$ in time, then the pair $(y(t),y(t+\pi/2))$ obeys a linear autonomous system.
\begin{equation}
\begin{cases}
\dot{y}(t)=y(t+\pi/2),\\[2pt]
\dfrac{\mathrm{d}}{\mathrm{d}t}\,y(t+\pi/2)=-y(t).
\end{cases}
\end{equation}

A simple way of reducing the augmented state~\eqref{eq:delay} is to apply to the stacked snapshot matrix $\mathbf{D}$ the same subspace (Grassmannian) framework used in POD, that is to treat the $N_f$ delayed copies as a single extended vector. The result, however, is that each mode becomes a space--time structure combining both spatial and temporal scales: one mode may contain both slow, large-scale and fast, small-scale features, and the temporal coherence of the underlying structures is thereby lost.

The way to proceed is to diagonalise the delay axis before carrying out the compression; for this purpose, we apply a windowed discrete Fourier transform (DFT) over the lag index,
\begin{equation}
\hat{\mathbf{D}}_k(t,\boldsymbol{\mu})=\sum_{j=0}^{N_f-1} w_j\, \mathbf{u}(t+j\,\delta t,\boldsymbol{\mu})\, e^{-2\pi \mathrm{i}\, jk/N_f},
\qquad k=0,\ldots,N_f-1,
\label{eq:dft_motivation}
\end{equation}
where $w_j$ is a tapering window. This process splits the augmented state into separate frequency components. We then perform an independent POD for each $\hat{\mathbf{D}}_k$. This is nothing other than Spectral Proper Orthogonal Decomposition (SPOD)~\cite{lumley_stochastic_1970, towne_spectral_2018}: it produces \emph{temporally coherent} modes, each linked to a single oscillation frequency, while at the same time maintaining POD optimality within each frequency band. As a result, the coefficients correspond to one coherent structure and one characteristic frequency each, {which can make their evolution smoother and easier to learn than that of conventional POD coefficients}. This insight is the basis of the approach presented below.

\subsection{Spectral Proper Orthogonal Decomposition}
\label{sec:spod}

We give a brief summary of SPOD in the formulation adopted here~\cite{towne_spectral_2018, schmidt_guide_2020}. Let $\mathbf{q}_n\in\mathbb{R}^{N}$, $n=0,\ldots,N_t-1$, denote a (mean-removed) vector-valued time series and let $\mathbf{G}\in\mathbb{R}^{N\times N}$ be a {symmetric positive-definite weighting matrix defining the inner product $\langle\mathbf{x},\mathbf{y}\rangle_{\mathbf G}=\mathbf{x}^{*}\mathbf G\mathbf y$ and the corresponding norm $\|\mathbf{x}\|_{\mathbf G}^2=\mathbf{x}^{*}\mathbf G\mathbf x$}. The data are divided into $N_b$ overlapping segments each containing $N_f$ samples (with $N_f$ odd), using a hop size $h=N_f-N_{\mathrm{ov}}$, which results in {$N_b=1+\lfloor(N_t-N_f)/h\rfloor$}. In this study a fully sliding approach is adopted with $h=1$ ($N_{\mathrm{ov}}=N_f-1$), which maximises the number of segments and gives one segment for each {admissible window position}, that is $N_b=N_t-N_f+1$. A Hamming window $w_j=0.54-0.46\cos\!\big(2\pi j/(N_f-1)\big)$ is applied to each segment, after which the segment is Fourier transformed,
\begin{equation}
\hat{\mathbf{q}}_k^{(b)}=\sum_{j=0}^{N_f-1} w_j\,\mathbf{q}_{bh+j}\,e^{-2\pi\mathrm{i}\,jk/N_f},\qquad
k=0,\ldots,\tfrac{N_f-1}{2},\quad b=0,\ldots,N_b-1,
\label{eq:block_dft}
\end{equation}
where, since the data are real-valued, only the non-negative frequencies $f_k=k/(N_f\,\delta t)$ are kept. At each frequency, the block realisations are then assembled into $\hat{\mathbf{Q}}_{k}=[\hat{\mathbf{q}}_k^{(0)},\ldots,\hat{\mathbf{q}}_k^{(N_b-1)}]\in\mathbb{C}^{N\times N_b}$. The SPOD modes at frequency $f_k$ are the eigenvectors of the weighted cross-spectral density $\hat{\mathbf{Q}}_k\hat{\mathbf{Q}}_k^{*}\mathbf{G}$; in practice, however, the snapshots method is used,
\begin{equation}
\hat{\mathbf{Q}}_{k}^{*}\mathbf{G}\,\hat{\mathbf{Q}}_{k}\,\boldsymbol{\Psi}_{k}=\boldsymbol{\Psi}_{k}\boldsymbol{\Lambda}_{k},
\label{eq:spod_eig}
\end{equation}
where $^{*}$ denotes the conjugate transpose, $\boldsymbol{\Psi}_k$ consists of the eigenvectors and $\boldsymbol{\Lambda}_k$ of the non-zero SPOD eigenvalues, and the modes are reconstructed as $\boldsymbol{\Phi}_{k}=\hat{\mathbf{Q}}_{k}\boldsymbol{\Psi}_{k}\boldsymbol{\Lambda}_{k}^{-1/2}$. An equivalent method, which we follow in our implementation, obtains $\boldsymbol{\Phi}_k$ by means of a truncated (randomised) singular value decomposition {of the weighted Fourier snapshots. More specifically, if $\mathbf{G}=\mathbf{L}_G\mathbf{L}_G^{*}$ is a Cholesky factorisation, we compute $\mathbf{L}_G^{*}\hat{\mathbf{Q}}_k=\mathbf{U}_k\boldsymbol\Sigma_k\mathbf{V}_k^{*}$ and recover the first $r$ modes as $\boldsymbol\Phi_k=\mathbf{L}_G^{-*}\mathbf{U}_{k,r}$}. By retaining the $r$ most important modes at each frequency, the SPOD coefficients are
\begin{equation}
\mathbf{a}_k^{(b)}=\boldsymbol{\Phi}_{k}^{*}\,\mathbf{G}\,\hat{\mathbf{q}}_k^{(b)}\in\mathbb{C}^{r},\qquad
\mathbf{A}_k=\boldsymbol{\Phi}_{k}^{*}\mathbf{G}\hat{\mathbf{Q}}_{k}=\boldsymbol{\Lambda}_{k}^{1/2}\boldsymbol{\Psi}_{k}^{*}.
\label{eq:spod_coefficients}
\end{equation}
In $\mathbf{A}_k$ each row relates to one SPOD mode and each column to a block; consequently, at any given frequency and mode one has a sequence of complex coefficients which show how the relevant coherent structure evolves from block to block. 

\paragraph{Inverse transform for the sliding window.}
When $h=1$, {each interior sample} appears in $N_f$ overlapping blocks; the sample at index $b+c$ is recovered by using the central point $c=(N_f-1)/2$ of block $b$, this being the point at which the Hamming window reaches its maximum ($w_c=1$) and where the reconstruction is affected as little as possible by frequency and rank truncation:
\begin{equation}
\mathbf{q}_{b+c}\approx \frac{1}{N_f\,w_c}\sum_{k\in\mathcal{K}} c_k\,\Re\!\left(\boldsymbol{\Phi}_k\mathbf{a}_k^{(b)}\,e^{2\pi\mathrm{i}\,kc/N_f}\right),
\label{eq:center_inverse}
\end{equation}
with $\mathcal{K}$ denoting the set of frequency bins that are retained, and $c_0=1$ and $c_k=2$ when $k\ge1$, these weights accounting for the contribution of the conjugate negative-frequency bins {(there is no Nyquist bin because $N_f$ is odd)}. Unlike the method of taking the last sample from each block, sampling at the centre avoids dividing by the small values of the tapered window near the ends and thus reduces the Gibbs-type oscillations at the boundaries between blocks.

\subsection{Overview of the SPOD-GPR framework}
\begin{figure}[t]
    \centering
    \includegraphics[width=\linewidth]{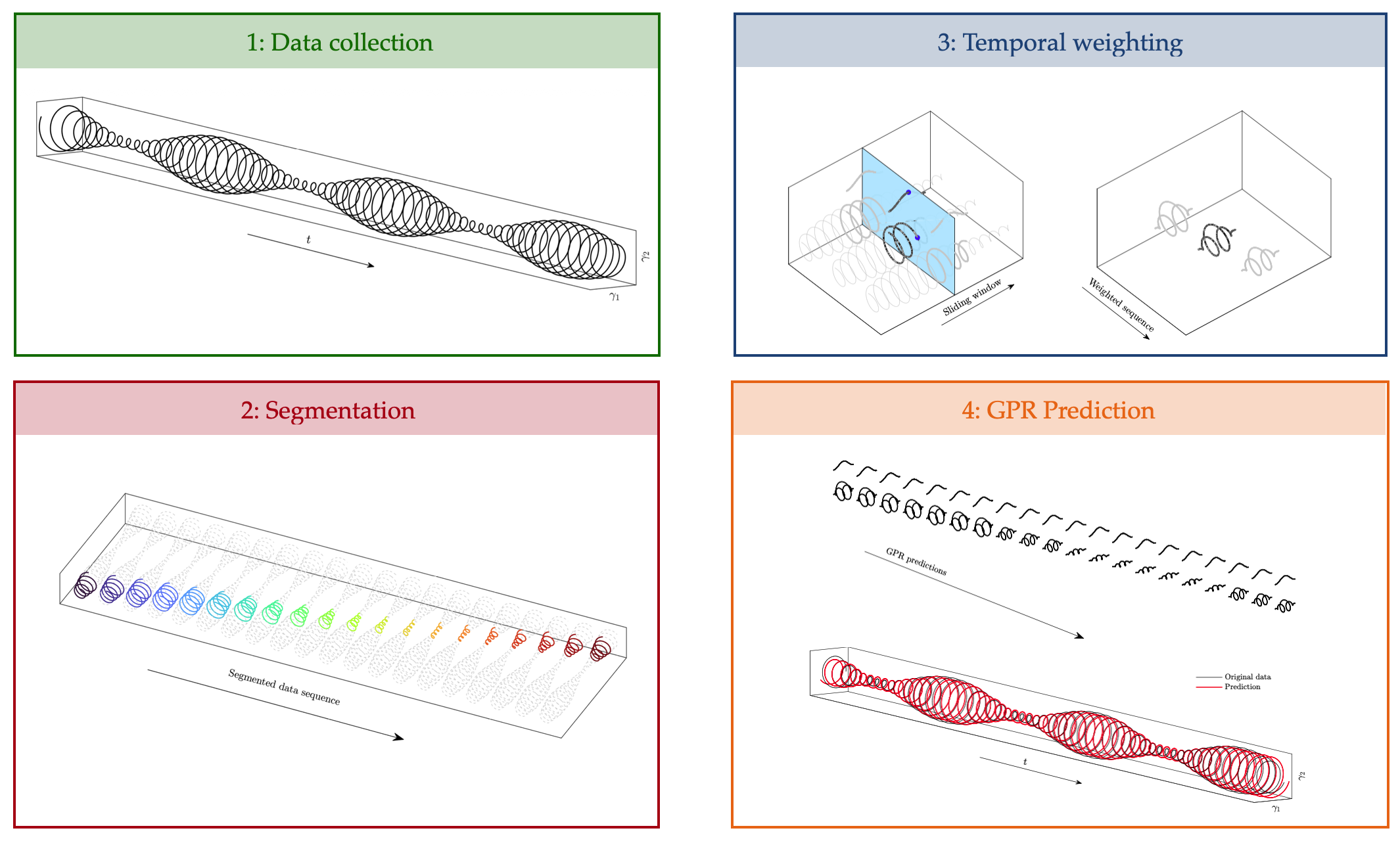}
    \caption{Sketch of the Gaussian Process Regression stage based on SPOD (when applied to the POD coordinates of stage~1). The high-fidelity data are obtained from the full-order model (1), divided over time into overlapping blocks (2), weighted using a temporal window and transformed into frequency space (3). GPR models are then trained on the resulting sequences of SPOD coefficients and are used to reconstruct the spatio-temporal dynamics (4).}
    \label{fig:spodgpr-pipeline}
\end{figure}

The proposed model approximates the parametric solution by
\begin{equation}
\mathbf{u}_n(\boldsymbol{\mu})\approx \bar{\mathbf{u}}(\boldsymbol{\mu})+\mathbf{L}_W^{-\top}\,\hat{\mathbf{V}}(\boldsymbol{\mu})\,\mathbf{q}_n(\boldsymbol{\mu}),
\qquad \mathbf W=\mathbf L_W\mathbf L_W^\top,
\label{eq:full_model}
\end{equation}
where the temporal mean is denoted by $\bar{\mathbf{u}}(\boldsymbol{\mu})$, $\hat{\mathbf{V}}(\boldsymbol{\mu})\in\mathbb{R}^{N_h\times R}$ is a parametric POD basis having {orthonormal columns in weighted coordinates}, and $\mathbf{q}_n(\boldsymbol{\mu})\in\mathbb{R}^R$ are the reduced coordinates. These coordinates are in turn represented by means of a parametric SPOD of rank $r$ for each of the $|\mathcal{K}|=K$ frequency bins. The approach is divided into two stages: in \emph{stage~1} a rank-$R$ parametric POD carries out the spatial compression (\Cref{sec:stage1}); in \emph{stage~2} a parametric SPOD-GPR is employed to model the temporal behaviour of the $R$ POD coordinates (\Cref{sec:stage2}). The reason for this two-stage procedure is computational efficiency: since blocking, Fourier transforms, and per-frequency SVDs depend on the spatial dimension, applying SPOD to $R\ll N_h$ coordinates allows the use of large time windows $N_f$ and thus enables a finer frequency resolution at a reasonable cost. As the POD coordinates are defined with respect to the $\mathbf{W}$-inner product, the reduced energy corresponds to the $\mathbf{W}$-weighted physical energy captured by the POD, and no further weighting is necessary in stage~2 ($\mathbf{G}=\mathbf{I}$). A summary of the SPOD-GPR step is given in \Cref{fig:spodgpr-pipeline}: the data obtained from the FOM (after having been reduced by POD) are divided into overlapping time blocks, multiplied by a moving temporal window, and transformed into the frequency domain; GPR models are then trained on the resulting SPOD coefficient sequences and are used to predict the spatio-temporal dynamics for new parameters and times.

\subsection{Stage 1: parametric POD}
\label{sec:stage1}

For every training parameter, {from the set $\mathcal P_{\mathrm{train}}=\{\boldsymbol\mu_i\}_{i=1}^{N_\mu}$}, we first evaluate the temporal average $\bar{\mathbf{u}}(\boldsymbol{\mu}_i)=N_t^{-1}\sum_n\mathbf{u}_n(\boldsymbol{\mu}_i)$ and then, with $\mathbf 1\in\mathbb R^{N_t}$, assemble the weighted fluctuation matrix
\begin{equation}
\mathbf{X}_i=\mathbf{L}_W^\top\big(\mathbf{U}_i-\bar{\mathbf{u}}(\boldsymbol{\mu}_i)\mathbf{1}^\top\big)\in\mathbb{R}^{N_h\times N_t},
\qquad \mathbf U_i=[\mathbf u_0(\boldsymbol\mu_i),\ldots,\mathbf u_{N_t-1}(\boldsymbol\mu_i)].
\end{equation}
The approximation $\mathbf{X}_i\approx\mathbf{V}_i\boldsymbol{\Theta}_i$ is obtained by applying a truncated randomized SVD, where $\mathbf{V}_i\in\mathbb{R}^{N_h\times R}$ has {orthonormal columns} and $\boldsymbol{\Theta}_i=\mathbf{V}_i^\top\mathbf{X}_i\in\mathbb{R}^{R\times N_t}$ contains the POD coefficients, one column per time instant. The rank $R$ is the same for all the parameters and, similarly to the common frequency set of \Cref{sec:stage2}, is selected from the energies averaged over the training set: denoting by $\sigma_j(\boldsymbol\mu_i)$ the singular values of $\mathbf X_i$, we compute $\bar E_j=N_\mu^{-1}\sum_i\sigma_j^2(\boldsymbol\mu_i)$ and take $R$ as the smallest rank for which $\sum_{j\le R}\bar E_j$ exceeds a prescribed fraction of the averaged total energy $N_\mu^{-1}\sum_i\|\mathbf X_i\|_F^2$ (alternatively, $R$ can be specified explicitly). It should be noted that a single global POD is not constructed; rather, each parameter is linked to its individual basis, which allows the modes to vary with $\boldsymbol{\mu}$.

Since an SVD determines its singular subspaces only up to an orthogonal change of basis (for example, due to sign ambiguities and rotations when the singular values are nearly identical), the unprocessed bases $\mathbf{V}_i$ are not smooth functions of $\boldsymbol{\mu}$. In order to ensure consistency, each basis is aligned with a common reference frame $\mathbf V_{\mathrm{ref}}$. This frame is a representative of the Karcher (Riemannian) mean of the training subspaces on the Grassmann manifold $\mathrm{Gr}(R,N_h)$~\cite{karcher_riemannian_1977,absil_riemannian_2004,amsallem_interpolation_2008}, that is, the minimiser of $\sum_i d^2\big(\operatorname{span}\mathbf V,\operatorname{span}\mathbf V_i\big)$, with $d$ the geodesic distance defined by the principal angles; it is computed by the standard fixed-point iteration that alternates the Grassmann logarithm and exponential maps. Placing the reference at the centre of the data keeps the principal angles to all the training subspaces small, and hence away from the singularity of the chart of \Cref{sec:grassmann}. The alignment is performed by means of an orthogonal Procrustes problem,
\begin{equation}
\mathbf{T}_i=\arg\min_{\mathbf{T}^\top\mathbf{T}=\mathbf{I}}\big\|\mathbf{V}_{\mathrm{ref}}-\mathbf{V}_i\mathbf{T}\big\|_F
=\mathbf{P}\mathbf{O}^\top,\qquad \mathbf{V}_i^\top\mathbf{V}_{\mathrm{ref}}=\mathbf{P}\boldsymbol{\Sigma}\mathbf{O}^\top,\label{eq:procrustes}
\end{equation}
which yields the rotated basis and coefficients $\hat{\mathbf{V}}_i=\mathbf{V}_i\mathbf{T}_i$ and $\hat{\boldsymbol{\Theta}}_i=\mathbf{T}_i^\top\boldsymbol{\Theta}_i$; these preserve the reconstruction, since $\hat{\mathbf{V}}_i\hat{\boldsymbol{\Theta}}_i=\mathbf{V}_i\boldsymbol{\Theta}_i$. For complex frames the same formula holds with $^\top$ replaced by $^*$. The mapping $\boldsymbol{\mu}\mapsto\hat{\mathbf{V}}(\boldsymbol{\mu})$ is modelled using the manifold-aware GPR as described in \Cref{sec:grassmann}, while the mean field $\boldsymbol{\mu}\mapsto\bar{\mathbf{u}}(\boldsymbol{\mu})$ is learned by means of a separate GP in physical space. It is important to note that the mean is not included in the POD coordinates, because if the two independently interpolated terms were multiplied together accuracy would be degraded between training points when the individual parameter-dependent bases contain low-energy directions that are specific to each parameter.

\subsection{Regression of orthonormal bases on a Grassmann chart}
\label{sec:grassmann}

Simply regressing the individual components of $\hat{\mathbf{V}}(\boldsymbol{\mu})$ would generally violate the orthonormality condition. Instead, we carry out the regression on coordinates that are derived from a local chart of the Grassmann manifold which is based at a reference basis~\cite{edelman_geometry_1998,jensen_canonically_2026}. Orthonormality is understood in the Euclidean inner product, since the physical weighting has already been absorbed ($\mathbf{L}_W^\top$ in stage~1, $\mathbf{G}=\mathbf{I}$ in stage~2). Let $\mathbf{B}=\mathbf{V}_{\mathrm{ref}}\in\mathbb{R}^{N_h\times R}$ (or $\mathbb{C}$ when considering SPOD modes) represent the reference frame, and let $\mathbf{F}=\hat{\mathbf{V}}_i$ be the aligned frame. We parametrise the subspace spanned by $\mathbf F$; its orientation within that subspace is fixed by the Procrustes alignment, as shown below. We define $\mathbf{C}=\mathbf{B}^{*}\mathbf{F}$ and
\begin{equation}
\mathbf{Y}=(\mathbf{I}-\mathbf{B}\mathbf{B}^{*})\,\mathbf{F}\,\mathbf{C}^{-1},
\end{equation}
so that the subspace associated with $\mathbf{F}$ can be expressed as the graph $\operatorname{span}(\mathbf{B}+\mathbf{Y})$ over the reference subspace; this gives rise to an affine chart on the Grassmann manifold and is well defined so long as $\mathbf{C}$ is nonsingular {(equivalently, none of the principal angles between the two subspaces equals $\pi/2$)}. {The Procrustes alignment fixes the in-subspace orientation but does not change this nonsingularity condition.} Since $\mathbf{B}^{*}\mathbf{Y}=\mathbf{0}$, the matrix $\mathbf{Y}$ has only $(N_h-R)R$ independent parameters. In order to determine these parameters without having to explicitly construct an orthogonal complement to $\mathbf{B}$, we select $R$ rows of $\mathbf B$ by means of the maxvol algorithm~\cite{olshevsky_maximal-volume_2001,olshevsky_how_2010}, initialised with partially pivoted Gaussian elimination; this gives a permutation $\boldsymbol{\Pi}$ such that the $R\times R$ subblock $\mathbf{B}_1$ in $\boldsymbol{\Pi}\mathbf{B}=[\mathbf{B}_1;\mathbf{B}_2]$ has (locally) maximal volume $|\det\mathbf B_1|$, which guarantees that all the entries of $\mathbf B_2\mathbf B_1^{-1}$ are bounded by one in modulus and hence that $\mathbf B_1$ is well conditioned; we then define $\mathbf{M}$ as $\mathbf{B}_2\mathbf{B}_1^{-1}$ and, upon writing $\boldsymbol{\Pi}\mathbf{Y}=[\mathbf{Y}_1;\mathbf{Y}_2]$, define
\begin{equation}
\boldsymbol{\Gamma}=\mathbf{Y}_2-\mathbf{M}\mathbf{Y}_1\in\mathbb{R}^{(N_h-R)\times R}.
\end{equation}
It is possible to verify that $\mathbf{Y}=(\mathbf{I}-\mathbf{B}\mathbf{B}^{*})\boldsymbol{\Pi}^\top[\mathbf{0};\boldsymbol{\Gamma}]$, this demonstrating that $\boldsymbol{\Gamma}$ gives a true and unconstrained parameterisation of the subspace.

\begin{figure}[t]
\centering
\begin{tikzpicture}[scale=0.92,
    lab/.style={font=\scriptsize},
    tag/.style={font=\tiny}]

\draw[thick] plot[domain=-3.2:3.2,smooth,samples=50] (\x,{-0.18*\x*\x});
\node[lab,anchor=west] at (3.25,{-0.18*3.2*3.2}) {$\mathrm{Gr}(R,N_h)$};

\draw[blue,thick] (-3.4,0)--(3.4,0);
\node[lab,blue,anchor=west] at (3.45,0) {$\mathbb{R}^{(N_h-R)\times R}$};

\fill[blue] (0,0) circle (1.8pt);
\node[lab,blue,anchor=south] at (0,0.05) {$\mathbf{B}$};

\foreach \x/\xt/\j/\anc/\dx in {-2.6/-2.99/1/north west/0.05,
                                 1.3/1.50/2/north east/-0.05,
                                 2.6/2.99/3/north east/-0.05}{
    \fill (\x,{-0.18*\x*\x}) circle (1.6pt);
    \node[lab,anchor=\anc] at ({\x+\dx},{-0.18*\x*\x-0.06}) {$\mathbf{F}_{\j}$};
    \draw[-{Stealth[length=3pt]},gray,dashed]
        (\x,{-0.18*\x*\x+0.06})--(\xt,-0.06);
    \fill[gray] (\xt,0) circle (1.3pt);
    \node[lab,anchor=south,gray!40!black] at (\xt,0.05) {$\boldsymbol{\Gamma}_{\j}$};
}
\node[tag,gray!40!black,anchor=west] at (2.88,-0.62) {chart};

\fill[red] (-1.84,0) circle (1.8pt);
\node[lab,red,anchor=south] at (-1.84,0.05) {$\boldsymbol{\Gamma}^\ast$};
\draw[-{Stealth[length=3.5pt]},red,thick]
    (-1.84,-0.06)--(-1.6,{-0.18*1.6*1.6+0.07});
\fill[red] (-1.6,{-0.18*1.6*1.6}) circle (1.8pt);
\node[lab,red,anchor=north west] at (-1.55,{-0.18*1.6*1.6-0.06}) {$\mathbf{F}^\ast$};
\node[tag,red,anchor=east] at (-1.78,-0.27) {\eqref{eq:synthesis}};

\end{tikzpicture}

\caption{Schematic of the regression of orthonormal frames on the Grassmann chart, in stage-1 notation (in stage~2, $\mathrm{Gr}(r,R)$ and $\boldsymbol\Gamma_k\in\mathbb{C}^{(R-r)\times r}$ for each bin $k$). The aligned training frames $\mathbf{F}_j=\hat{\mathbf V}(\boldsymbol\mu_j)$ are mapped to unconstrained coordinates $\boldsymbol\Gamma_j$ relative to the reference frame $\mathbf B$, which corresponds to $\boldsymbol\Gamma=\mathbf 0$. The GP predicts $\boldsymbol\Gamma^\ast=\boldsymbol\Gamma(\boldsymbol\mu^\ast)$, which~\eqref{eq:synthesis} maps back to an exactly orthonormal frame $\mathbf{F}^\ast$, Procrustes-aligned to $\mathbf B$. Every point of the coordinate space corresponds to a valid frame.}
\label{fig:grassmann_interpolation}
\end{figure}
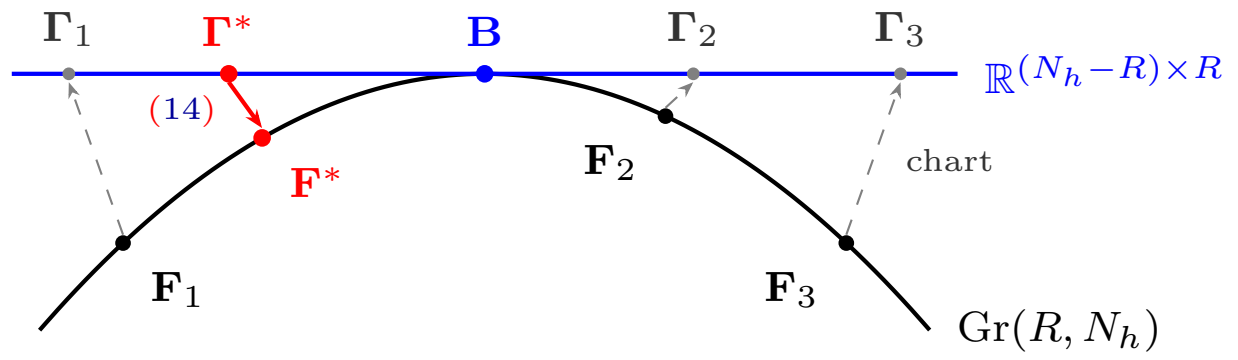

No additional coordinate is needed for the orientation of the frame within its subspace, because it is fixed by the Procrustes alignment. Indeed, by~\eqref{eq:procrustes}, the aligned frame satisfies $\mathbf{C}=\mathbf{B}^{*}\mathbf{F}=\mathbf{O}\boldsymbol{\Sigma}\mathbf{O}^{*}$, which is Hermitian positive definite; since $\mathbf{F}=(\mathbf{B}+\mathbf{Y})\mathbf{C}$ and $\mathbf{F}^{*}\mathbf{F}=\mathbf{I}$, it follows that $\mathbf{I}+\mathbf{Y}^{*}\mathbf{Y}=\mathbf{C}^{-2}$, i.e.\ $(\mathbf{I}+\mathbf{Y}^{*}\mathbf{Y})^{-1/2}=\mathbf{C}$. From the predicted values of $\boldsymbol{\Gamma}$, the frame is therefore reconstructed as

\begin{equation}
\mathbf{F}=(\mathbf{B}+\mathbf{Y})\,(\mathbf{I}+\mathbf{Y}^{*}\mathbf{Y})^{-1/2},\qquad
\mathbf{Y}=(\mathbf{I}-\mathbf{B}\mathbf{B}^{*})\boldsymbol{\Pi}^\top\begin{bmatrix}\mathbf{0}\\ \boldsymbol{\Gamma}\end{bmatrix}.
\label{eq:synthesis}
\end{equation}
Conversely, for any $\boldsymbol{\Gamma}$ this construction yields an exactly orthonormal frame ($\mathbf{F}^{*}\mathbf{F}=\mathbf{I}$) with $\mathbf{B}^{*}\mathbf{F}=(\mathbf{I}+\mathbf{Y}^{*}\mathbf{Y})^{-1/2}$ Hermitian positive definite, i.e.\ a frame that is Procrustes-aligned to $\mathbf{B}$. Every predicted or sampled basis is therefore a valid orthonormal frame, expressed in the same aligned convention as the training frames and the rotated modal coefficients, so that predicted modes and coefficients remain consistent. This requires the Procrustes reference to coincide with the base of the chart and no further rotation to be applied after the alignment. Since $\boldsymbol{\Gamma}$ lies in a vector space, its entries can be regressed directly. The procedure is sketched in \Cref{fig:grassmann_interpolation}.

\paragraph{Specialisation.} In stage~1 the chart is applied with real frames, $\mathbf{B}=\mathbf{V}_{\mathrm{ref}}\in\mathbb{R}^{N_h\times R}$ and $\mathbf{F}=\hat{\mathbf{V}}_i$, so that $\boldsymbol\Gamma\in\mathbb{R}^{(N_h-R)\times R}$. In stage~2 it is applied separately at each retained bin $k\in\mathcal K$ with complex frames $\mathbf{B}_k=\boldsymbol\Phi_{k,\mathrm{ref}}\in\mathbb{C}^{R\times r}$, the Karcher-mean reference of bin $k$ (\Cref{sec:stage2}), and $\mathbf{F}_k=\hat{\boldsymbol\Phi}_k(\boldsymbol\mu_i)$, i.e.\ with $N_h\to R$ and $R\to r$, so that $\boldsymbol\Gamma_k\in\mathbb{C}^{(R-r)\times r}$, whose real and imaginary parts are fitted separately. The cost is $\mathcal{O}(N_hR^2)$ per frame in stage~1 and $\mathcal{O}(Rr^2)$ per frame and bin in stage~2.

\subsection{Stage 2: parametric SPOD-GPR on the POD coordinates}
\label{sec:stage2}

The aligned, zero-mean POD coordinates
$\hat{\boldsymbol{\Theta}}_i=[\mathbf{q}_0(\boldsymbol{\mu}_i),\ldots,\mathbf{q}_{N_t-1}(\boldsymbol{\mu}_i)]\in\mathbb{R}^{R\times N_t}$
are used as input in stage~2, i.e. $\mathbf{q}_n(\boldsymbol{\mu}_i)$ is the $n$-th column of $\hat{\boldsymbol{\Theta}}_i$; they form the time series of \Cref{sec:spod} with $N=R$ and $\mathbf{G}=\mathbf{I}$.

\paragraph{Common frequency set.} For each training parameter we calculate the Parseval-weighted DFT energy in each bin, that is $E_k(\boldsymbol{\mu}_i)=c_k\sum_b\|\hat{\mathbf{q}}_k^{(b)}(\boldsymbol{\mu}_i)\|^2$, then average the energies over the training set and choose the $K$ bins having the highest energy, calling this set $\mathcal{K}$. This common set of bins is used for all the parameters, so there is no need to interpolate the frequencies in $\boldsymbol{\mu}$ and the computations can be limited to the selected bins (by means of a direct DFT).

\paragraph{Per-parameter SPOD and alignment.} For each training parameter, a separate rank-$r$ SPOD is carried out on the bins in $\mathcal{K}$, resulting in complex modes $\boldsymbol{\Phi}_k(\boldsymbol{\mu}_i)\in\mathbb{C}^{R\times r}$ and coefficients $\mathbf{a}_k^{(b)}(\boldsymbol{\mu}_i)$. At each bin, a reference frame $\boldsymbol\Phi_{k,\mathrm{ref}}$ is computed, as in stage~1, as a representative of the Karcher mean of the training subspaces $\{\operatorname{span}\boldsymbol\Phi_k(\boldsymbol\mu_i)\}_{i=1}^{N_\mu}$, here on the complex Grassmann manifold $\mathrm{Gr}(r,\mathbb{C}^R)$~\cite{karcher_riemannian_1977,absil_riemannian_2004}. The modes corresponding to each parameter are then rotated so as to coincide with $\boldsymbol\Phi_{k,\mathrm{ref}}$ by solving the \emph{unitary} Procrustes problem, that is, the complex version of~\eqref{eq:procrustes}; the coefficients are then rotated accordingly, {$\hat{\boldsymbol\Phi}_{k,i}=\boldsymbol\Phi_{k,i}\mathbf T_{k,i}$ and $\hat{\mathbf a}_{k,i}^{(b)}=\mathbf T_{k,i}^{*}\mathbf a_{k,i}^{(b)}$, with $\mathbf T_{k,i}$ unitary}, this rotation preserving the reconstruction {$\hat{\boldsymbol\Phi}_{k,i}\hat{\mathbf a}_{k,i}^{(b)}=\boldsymbol\Phi_{k,i}\mathbf a_{k,i}^{(b)}$}. No further rotation (for instance, an alignment across frequencies) is applied after this step, as required in \Cref{sec:grassmann}. The aligned modes are finally transformed, on a bin-by-bin basis, into $\boldsymbol{\Gamma}_k$ coordinates as explained in \Cref{sec:grassmann}. {For each retained frequency, the reference and aligned frames used in that construction are $\mathbf B_k=\boldsymbol\Phi_{k,\mathrm{ref}}$ and $\mathbf F_k=\hat{\boldsymbol\Phi}_k(\boldsymbol\mu_i)$, respectively. The corresponding chart coordinates are therefore $\boldsymbol\Gamma_k(\boldsymbol\mu_i)$, and the reconstruction~\eqref{eq:synthesis} is applied separately at each $k\in\mathcal K$.}

\begin{figure}[t]
\centering
\begin{tikzpicture}[
    font=\small, >={Stealth[length=4pt]},
    gp/.style={draw=acc, fill=acc!8, thick, rounded corners=3pt, minimum width=4.7cm, align=center, inner sep=4pt},
    dat/.style={draw=black!60, rounded corners=2pt, align=center, inner sep=3pt, minimum height=0.8cm},
    res/.style={draw=cR, fill=cR!8, thick, rounded corners=3pt, align=center, inner sep=4pt, minimum height=0.9cm},
    lab/.style={font=\scriptsize, align=center}]

\node[dat, minimum width=0.9cm] (mu) at (0,0) {$\boldsymbol\mu^\ast$};

\node[gp] (g1) at (3.9,2.3) {\textbf{SPOD-basis GP}\\
    $\boldsymbol\mu\mapsto\{\boldsymbol\Gamma_k(\boldsymbol\mu)\}_{k\in\mathcal K}$\\
    {\scriptsize\color{black!55}data: SPOD modes of each $\boldsymbol\mu_i$}};
\node[gp] (g2) at (3.9,0) {\textbf{Trajectory GP}\\
    $\boldsymbol\mu\mapsto[\mathbf z^{(0)},\ldots,\mathbf z^{(N_b-1)}](\boldsymbol\mu)$\\
    {\scriptsize\color{black!55}data: trajectory of each $\boldsymbol\mu_i$}};
\node[gp] (g3) at (3.9,-2.3) {\textbf{Dynamics GP}\\
    $(\mathbf z^{(b)},\boldsymbol\mu)\mapsto\Delta\mathbf z^{(b)}$\\
    {\scriptsize\color{black!55}data: increments over all $i$ and $b$}};

\draw[dashed, black!45, rounded corners=4pt] (1.35,3.25) rectangle (6.45,-3.35);
\node[lab, black!60] at (3.9,-3.62) {joint pathwise (Matheron) samples, $p=0,\ldots,N_s-1$};

\node[dat] (phi) at (9.1,2.3) {$\boldsymbol\Phi_k(\boldsymbol\mu^\ast)$, $k\in\mathcal K$};
\node[dat] (traj) at (9.1,0) {$\mathbf z^{(b)}(\boldsymbol\mu^\ast)$, $b\le N_b-1$};
\node[dat] (roll) at (9.1,-2.3) {$\mathbf z^{(b+1)}=\mathbf z^{(b)}+\Delta\mathbf z^{(b)}$\\[-1pt]
    {\scriptsize $b\ge N_b-1$ (autoregressive)}};

\node[dat] (cinv) at (13.7,0) {central inverse\\ \eqref{eq:center_inverse}};
\node[res] (u) at (13.7,-2.3) {$\mathbf u_n(\boldsymbol\mu^\ast)$ via \eqref{eq:full_model}\\
    {\scriptsize with $\bar{\mathbf u}$, $\hat{\mathbf V}$ from stage~1}};

\draw[->] (mu.east) -- (g2.west);
\draw[->] (mu.north) |- (g1.west);
\draw[->] (mu.south) |- (g3.west);
\draw[->] (g1.east) -- node[lab, above]{\eqref{eq:synthesis}} (phi.west);
\draw[->] (g2.east) -- (traj.west);
\draw[->] (traj.south) -- node[lab, left]{initial condition\\ $\mathbf z^{(N_b-1)}(\boldsymbol\mu^\ast)$} (roll.north);
\draw[->] ([yshift=4pt]roll.west) -- node[lab, above]{$\mathbf z^{(b)}$} ([yshift=4pt]g3.east);
\draw[->] ([yshift=-4pt]g3.east) -- node[lab, below]{$\Delta\mathbf z^{(b)}$} ([yshift=-4pt]roll.west);
\draw[->] (phi.east) -| (cinv.north);
\draw[->] (roll.east) -- ++(0.45,0) |- node[lab, pos=0.25, right]{$\hat{\mathbf a}_k^{(b)}$} ([yshift=-5pt]cinv.west);
\draw[->, dashed] ([yshift=5pt]traj.east) -- node[lab, above]{in-window} ([yshift=5pt]cinv.west);
\draw[->] (cinv.south) -- node[lab, right]{$\mathbf q_n(\boldsymbol\mu^\ast)$} (u.north);
\end{tikzpicture}
\caption{{Sketch of the three stage-2 Gaussian process surrogates and of their combination at prediction time (\Cref{sec:prediction}). For a query parameter $\boldsymbol\mu^\ast$, the SPOD-basis GP predicts the chart coordinates of the SPOD modes, which are mapped back to orthonormal modes $\boldsymbol\Phi_k(\boldsymbol\mu^\ast)$ by~\eqref{eq:synthesis}; the trajectory GP predicts the spectral coefficients over the training window, and its last block initialises the autoregressive rollout of the dynamics GP. Modes and coefficients are combined through the central inverse~\eqref{eq:center_inverse}, and the physical field is then reconstructed with the stage-1 quantities via~\eqref{eq:full_model}. The dashed arrow indicates the in-window reconstruction obtained from the trajectory GP alone. All the GPs are sampled jointly, in lock-step, by means of Matheron's rule.}}
\label{fig:stage2-gps}
\end{figure}
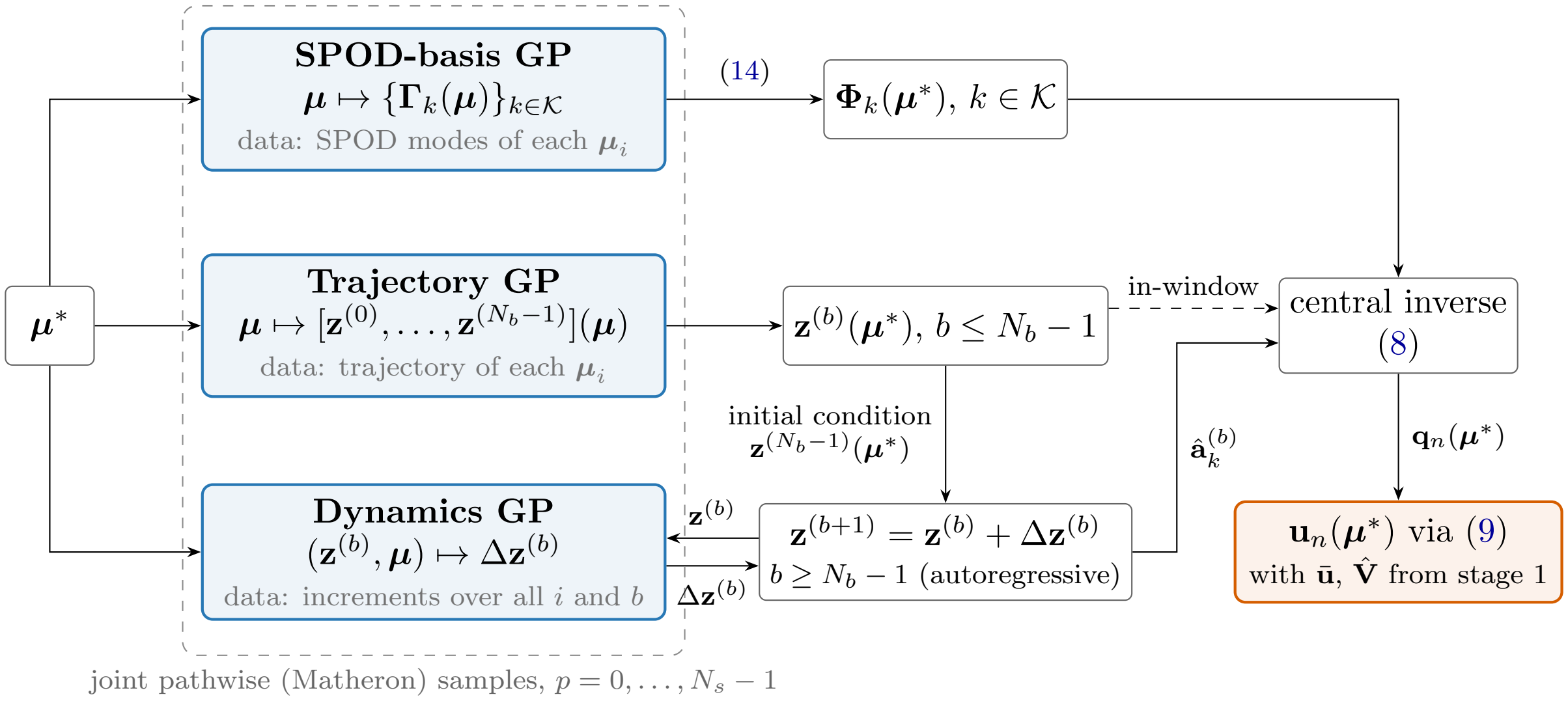

\paragraph{Three surrogates.} We define $\mathbf{z}^{(b)}(\boldsymbol{\mu})\in\mathbb{R}^{2rK}$ to be the vector consisting of the real and imaginary parts of $\{\hat{\mathbf{a}}_k^{(b)}(\boldsymbol{\mu})\}_{k\in\mathcal{K}}$. {Writing $\mathcal K=\{k_1,\ldots,k_K\}$, we fix the ordering}
\begin{equation}
\mathbf z^{(b)}(\boldsymbol\mu)=
\begin{bmatrix}
\Re\hat{\mathbf a}_{k_1}^{(b)}(\boldsymbol\mu)\\
\Im\hat{\mathbf a}_{k_1}^{(b)}(\boldsymbol\mu)\\
\vdots\\
\Re\hat{\mathbf a}_{k_K}^{(b)}(\boldsymbol\mu)\\
\Im\hat{\mathbf a}_{k_K}^{(b)}(\boldsymbol\mu)
\end{bmatrix}\in\mathbb R^{2rK}.
\end{equation}
We train three Gaussian process surrogates:
\begin{enumerate}
\item the \emph{{SPOD-basis GP}} $\boldsymbol{\mu}\mapsto\{\boldsymbol{\Gamma}_k(\boldsymbol{\mu})\}_{k\in\mathcal{K}}$, a single multi-output GP with $2K(R-r)r$ real outputs, whose outputs are mapped back to orthonormal SPOD modes using~\eqref{eq:synthesis} {with the frequency-specific reference frame $\mathbf B_k$};
\item the \emph{trajectory GP} $\boldsymbol{\mu}\mapsto[\mathbf{z}^{(0)}(\boldsymbol{\mu}),\ldots,\mathbf{z}^{(N_b-1)}(\boldsymbol{\mu})]$, which jointly predicts the complete coefficient trajectory over the training window, yielding a reconstruction that is truly parameter-dependent within that window;
\item the \emph{{dynamics GP}} $(\mathbf{z}^{(b)},\boldsymbol{\mu})\mapsto\Delta\mathbf{z}^{(b)}=\mathbf{z}^{(b+1)}-\mathbf{z}^{(b)}$, which is trained using pairs that have been aggregated across all training parameters and blocks and serves to establish a single evolution rule that is conditioned on the parameters, being employed thereafter for extrapolation beyond the training window.
\end{enumerate}
In the {dynamics} GP the parameter appears only as an input feature. By predicting the increment rather than the next state, the learning target is kept smaller and more nearly stationary when the trajectory changes smoothly from one block to the next, as is the case with the sliding window $h=1$. {During extrapolation, this GP is used autoregressively.} {The three surrogates and the way in which they are combined at prediction time are sketched in \Cref{fig:stage2-gps}.}

\subsection{Gaussian process regression}
\label{sec:gpr}

Each of the surrogate models (that is, the POD-mode GP and the stage~1 mean GP, along with the three stage~2 GPs) makes use of the same regression backend. The inputs are transformed onto the unit hypercube and the outputs are standardised. A Mat\'ern-$9/2$ {unit-variance correlation kernel $k_{\ell}(\mathbf{x},\mathbf{x}')$} is used, this function having {a length scale} $\ell=\gamma\,h_{\mathrm{fill}}$, where $h_{\mathrm{fill}}$ is the fill distance of a quasi-uniform set of inducing points obtained by farthest-point sampling within a ball that contains the inputs, and $\gamma$ is a constant that is specified by the user. The resulting model is
\begin{equation}
f\sim\mathcal{GP}(0,\sigma_f^2 k_{\ell}),\qquad y=f(\mathbf{x})+e,\quad e\sim\mathcal{N}(0,\sigma_f^2\lambda),
\end{equation}
and its posterior mean is identical to the kernel ridge regression predictor~\cite{kanagawa_gaussian_2018}. In order to achieve scalability when dealing with large data sets (in particular for the {dynamics} GP, which involves $N_\mu(N_b-1)$ observations), we approximate the kernel using Nystr\"om features~\cite{williams_using_2000}. More specifically, by selecting landmarks from (a subset of) the training samples together with the inducing points, we can write $k_{\ell}(\mathbf{x},\mathbf{x}')\approx\boldsymbol{\varphi}(\mathbf{x})^\top\boldsymbol{\varphi}(\mathbf{x}')$, thus preventing any $n\times n$ matrix from being formed. If we denote the feature matrix by {$\mathbf F_X=[\boldsymbol\varphi(\mathbf x_1),\ldots,\boldsymbol\varphi(\mathbf x_n)]^\top$}, then the posterior-mean weights are given by {$\boldsymbol{\mu}_w=(\mathbf F_X^\top\mathbf F_X+\lambda\mathbf{I})^{-1}\mathbf F_X^\top\mathbf{y}$}.

\paragraph{Pathwise sampling.} The posterior draws are obtained by using Matheron’s rule (pathwise conditioning)~\cite{wilson_efficiently_2020},
\begin{equation}
\mathbf{w}^{(p)}=\boldsymbol{\mu}_w+\sigma_f\left[\mathbf{w}_0^{(p)}-(\mathbf F_X^\top\mathbf F_X+\lambda\mathbf{I})^{-1}\mathbf F_X^\top(\mathbf F_X\mathbf{w}_0^{(p)}+\mathbf{e}^{(p)})\right],\quad
\mathbf{w}_0^{(p)}\sim\mathcal{N}(\mathbf{0},\mathbf{I}),\ \mathbf{e}^{(p)}\sim\mathcal{N}(\mathbf{0},\lambda\mathbf{I}),
\end{equation}
with $f^{(p)}(\mathbf{x}) = \boldsymbol{\varphi}(\mathbf{x})^\top \mathbf{w}^{(p)}$. Since each draw is an entire function, it can be evaluated consistently at a number of different input points. This feature is important for the {autoregressive} rollout, since the output of each step is used as the input for the following step. We sample $N_s$ functions from all of the GPs in a synchronized (lock-step) way; specifically, sample $0$ has {$\mathbf{w}_0^{(0)} = \mathbf{e}^{(0)} = \mathbf{0}$}, and therefore matches the posterior mean for each GP. The draw simplifies: with $\mathbf H=\mathbf F_X^\top\mathbf F_X+\lambda\mathbf I$, one has $\mathbf w_0^{(p)}-\mathbf H^{-1}\mathbf F_X^\top\mathbf F_X\mathbf w_0^{(p)}=\lambda\mathbf H^{-1}\mathbf w_0^{(p)}$, hence $\mathbf w^{(p)}=\boldsymbol\mu_w+\sigma_f\mathbf H^{-1}\big(\lambda\mathbf w_0^{(p)}-\mathbf F_X^\top\mathbf e^{(p)}\big)$, whose perturbation is distributed as $\mathcal N(\mathbf 0,\sigma_f^2\lambda\mathbf H^{-1})$, i.e.\ the exact weight-space posterior. We therefore draw $\mathbf w^{(p)}=\boldsymbol\mu_w+\sigma_f\sqrt{\lambda}\,\mathbf L_H^{-\top}\boldsymbol\eta^{(p)}$, with $\mathbf H=\mathbf L_H\mathbf L_H^\top$, $\boldsymbol\eta^{(p)}\sim\mathcal N(\mathbf 0,\mathbf I)$ and $\boldsymbol\eta^{(0)}=\mathbf 0$, so that a single factorisation of $\mathbf H$, whose size is the number of Nystr\"om features, serves all outputs and samples.

\paragraph{Calibration of the variance.} The variance hyperparameter {$\sigma_f^2$} has no effect on the posterior mean and is therefore adjusted separately for each output using the closed-form leave-one-out (LOO) residuals. If we let {$\mathbf{C}_\lambda=(\mathbf F_X\mathbf F_X^\top+\lambda\mathbf{I})^{-1}$} and $\boldsymbol{\alpha}=\mathbf{C}_\lambda\mathbf{y}$, the LOO residual at point $i$ is given by {$r_i=\alpha_i/(\mathbf C_\lambda)_{ii}$} and has a unit-scale standard deviation of {$(\mathbf C_\lambda)_{ii}^{-1/2}$}. By the Woodbury identity, $\mathbf C_\lambda=\lambda^{-1}\big(\mathbf I-\mathbf F_X\mathbf H^{-1}\mathbf F_X^\top\big)$, so that $\boldsymbol\alpha=\lambda^{-1}(\mathbf y-\mathbf F_X\boldsymbol\mu_w)$ and $(\mathbf C_\lambda)_{ii}=\lambda^{-1}\big(1-\boldsymbol\varphi(\mathbf x_i)^\top\mathbf H^{-1}\boldsymbol\varphi(\mathbf x_i)\big)$; both are computed from the factorisation of $\mathbf H$, and no $n\times n$ matrix is formed. The value of {$\sigma_f$} is selected so that the nominal $95\%$ interval includes $95\%$ of the standardised LOO residuals. In the case of the {dynamics} GP, it is too optimistic to omit just one step since adjacent blocks from the same trajectory are almost identical; instead, entire training trajectories are left out (leave-one-parameter-out), as this more accurately represents prediction for an unseen parameter. {For this last model, the variance calibration is therefore based on the leave-one-parameter-out residuals rather than the pointwise LOO residuals.}

\subsection{Prediction}
\label{sec:prediction}

For a given query parameter $\boldsymbol{\mu}^\ast$ and a prediction horizon of $N_{\mathrm{test}}$ steps beyond the training interval, and for each sample {$p=0,\ldots,N_s-1$}:
\begin{enumerate}
\item the trajectory GP is queried at $\boldsymbol{\mu}^\ast$, and its final block $\mathbf{z}^{(N_b-1)}_p(\boldsymbol{\mu}^\ast)$ is taken as the parameter-dependent initial condition;
\item the {dynamics} GP is propagated according to $\mathbf{z}^{(b+1)}_p=\mathbf{z}^{(b)}_p+\Delta\mathbf{z}_p(\mathbf{z}^{(b)}_p,\boldsymbol{\mu}^\ast)$ for $N_{\mathrm{test}}+(N_f-1-c)$ blocks, where the additional blocks offset the delay induced by the central inverse~\eqref{eq:center_inverse};
\item the SPOD modes $\boldsymbol{\Phi}_{k,p}(\boldsymbol{\mu}^\ast)$ are drawn from the {SPOD-basis} GP and then decoded using~\eqref{eq:synthesis}, while the POD coordinates are recovered via~\eqref{eq:center_inverse};
\item the POD basis $\hat{\mathbf{V}}_p(\boldsymbol{\mu}^\ast)$ and the mean $\bar{\mathbf{u}}_p(\boldsymbol{\mu}^\ast)$ are sampled from the stage-1 GPs and the complete field is then reconstructed using~\eqref{eq:full_model}.
\end{enumerate}
The estimate is given by sample $0$ (the posterior mean {in the latent GP outputs, followed by nonlinear reconstruction}), and the pointwise predictive variance is obtained as {the empirical mean squared deviation of samples $1,\ldots,N_s-1$ from sample $0$}. As a result, all the sources of uncertainty (the mean field, the POD basis, the SPOD modes, the initial condition, and the dynamics) are propagated together and in a nonlinear manner into the physical field. If the same method is applied but steps~1--2 are omitted—that is, if the trajectory GP is used for each block—a parameter-dependent reconstruction is obtained over the training interval.

\subsection{Baseline: POD-GPR}
\label{sec:baseline}

As a reference, we consider a parametric POD-GPR that employs exactly the same regression pipeline: for each parameter, the weighted POD bases of rank $r_{\mathrm{POD}}$ are aligned by means of a Procrustes analysis and then mapped or regressed onto a Grassmann chart, after which a mean GP, a trajectory GP defined by {$\boldsymbol{\mu}\mapsto[\boldsymbol\xi_0(\boldsymbol{\mu}),\ldots,\boldsymbol\xi_{N_t-1}(\boldsymbol{\mu})]$}, and a {dynamics} GP given by {$(\boldsymbol\xi_n,\boldsymbol\mu)\mapsto\Delta\boldsymbol\xi_n=\boldsymbol\xi_{n+1}-\boldsymbol\xi_n$} are used, with the same kernels, calibration procedure, and Matheron sampling. {Here $\boldsymbol\xi_n\in\mathbb R^{r_{\mathrm{POD}}}$, the $n$-th column of the aligned rank-$r_{\mathrm{POD}}$ coefficient matrix, denotes the POD reduced state of the baseline, distinct from the spectral latent state $\mathbf z^{(b)}$ of the proposed method.} For a query parameter $\boldsymbol\mu^\ast$, the baseline prediction reads
\begin{equation}
\mathbf{u}_n(\boldsymbol{\mu}^\ast)\approx \bar{\mathbf{u}}(\boldsymbol{\mu}^\ast)+\mathbf{L}_W^{-\top}\,\hat{\mathbf{V}}_{\mathrm{POD}}(\boldsymbol{\mu}^\ast)\,\boldsymbol\xi_n(\boldsymbol{\mu}^\ast),
\qquad
\boldsymbol\xi_{n+1}=\boldsymbol\xi_n+\Delta\boldsymbol\xi(\boldsymbol\xi_n,\boldsymbol\mu^\ast),\quad n\ge N_t-1,
\label{eq:baseline}
\end{equation}
where $\hat{\mathbf V}_{\mathrm{POD}}(\boldsymbol\mu^\ast)\in\mathbb R^{N_h\times r_{\mathrm{POD}}}$ is reconstructed from the chart via~\eqref{eq:synthesis}, and the rollout starts from the last training state $\boldsymbol\xi_{N_t-1}(\boldsymbol\mu^\ast)$ given by the trajectory GP. Compared with~\eqref{eq:full_model}, the reduced coordinates are propagated directly in time, one snapshot per step, instead of through the windowed SPOD blocks and the central inverse~\eqref{eq:center_inverse}. The only difference from SPOD-GPR is the choice of spectral latent representation.

\section{Numerical experiments}
\label{sec:results}

\subsection{Set-up and error measures}

In every experiment the parameters are selected uniformly across their respective ranges. For each simulation the first $95\%$ of the snapshots (with rounding so that the blocking procedure is exact) are used for training and the remaining ones for testing. Since the test parameters are not seen during training, each test involves both parametric interpolation and temporal extrapolation. We apply SPOD with a window of $N_f=101$, a stride of $h=1$, and a Hamming taper; a stage-1 POD rank of $R=50$; and GPs with a Mat\'ern-$9/2$ kernel, a ridge value of $\lambda=10^{-12}$, 50 inducing points for the dynamics GP, and $N_s=20$ Matheron samples. The case-specific hyperparameters are provided in \Cref{tab:hyper}. In all the cases, the truncation of both methods (the $r_{\mathrm{POD}}$ POD modes of POD-GPR, and the $K$ frequency bins with $r$ modes each of SPOD-GPR) preserves approximately $95\%$ of the training energy.
Let $u$ and $\tilde u$ stand for the actual and predicted test arrays over the test times, degrees of freedom, and test parameters; let $\mathcal{R}=\max u-\min u$ be the overall range of $u$, and let $v$ represent the predictive variance. We give the normalised mean absolute error (NMAE) and the normalised root-mean-square error (NRMSE),
\begin{equation}
\mathrm{NMAE}=\frac{\operatorname{mean}|\tilde u-u|}{\mathcal{R}},\qquad
\mathrm{NRMSE}=\frac{\sqrt{\operatorname{mean}(\tilde u-u)^2}}{\mathcal{R}},
\end{equation}
as well as the normalised predicted standard deviation $\mathrm{NSD}=\sqrt{\operatorname{mean} v}/\mathcal{R}$, which for a well-calibrated model should be directly comparable to the NRMSE {; their ratio NSD/NRMSE (spread--skill ratio) is therefore close to one for a calibrated model, and smaller (larger) than one for an overconfident (conservative) model}. We also give the normalised $95\%$ interval half-width $1.96\,\mathrm{NSD}$ and the empirical coverage $\mathrm{COV}$, defined as the proportion of entries for which $|\tilde u-u|\le 1.96\sqrt{v}$. The time histories are obtained by averaging only over the degrees of freedom and the test parameters, while the spatial maps are derived by averaging over time and the test parameters. Furthermore, we provide the spatial average of $\overline{|u|}$, that is, the mean over the test parameters of the pointwise magnitude, as an indication of the amplitude of the predicted fields.

\begin{table}[t]
\centering
\caption{Case-dependent settings. $r$ is the number of SPOD modes per frequency (SPOD-GPR) and $r_{\mathrm{POD}}$ the number of POD modes (POD-GPR); $K$ is the number of frequency bins retained; $\gamma$ is the length-scale coefficient of all the GPs; $N_\mu^{\mathrm{train}}/N_\mu^{\mathrm{test}}$ are the numbers of training and test parameters{, and $N_t^{\mathrm{train}}/N_{\mathrm{test}}$ the numbers of training and test snapshots for each parameter}.}
\label{tab:hyper}
\begin{tabular}{lccccccc}
\toprule
Case & parameter $\boldsymbol{\mu}$ & $N_\mu^{\mathrm{train}}/N_\mu^{\mathrm{test}}$ & $N_t^{\mathrm{train}}/N_{\mathrm{test}}$ & $r$ & $K$ & $r_{\mathrm{POD}}$ & $\gamma$\\
\midrule
KS, periodic & $\nu_1\in[0.45,0.55]$ & 80/20 & 950/50 & 1 & 3 & 2 & 5\\
KS, travelling wave & $\nu_2\in[0.3,0.4]$ & 80/20 & 950/50 & 4 & 20 & 6 & 5\\
Cylinder, Reynolds & $Re\in[110,160]$ & 80/20 & 950/50 & 2 & 3 & 2 & 50\\
Elliptic obstacle & $(a,b)\in[0.45,0.55]^2$ & 80/20 & 950/50 & 2 & 3 & 2 & 50\\
\bottomrule
\end{tabular}
\end{table}

The results for the four cases are summarised in \Cref{tab:results} and discussed below.

\begin{table}[t]
\centering
\caption{Space--time averaged results for the test window (unseen parameters and future times). HW95: the normalised $95\%$ half-width $1.96\,\mathrm{NSD}$; COV: the empirical coverage of the $95\%$ interval; $\langle\overline{|u|}\rangle$: the space--time average of the mean magnitude {; NSD/NRMSE: the spread--skill ratio, close to one for a calibrated model}. The best errors are given in bold. }
\label{tab:results}
\small
\setlength{\tabcolsep}{4pt}
\begin{tabular}{llcccccc}
\toprule
Case & Method & NMAE & NRMSE & HW95 & {NSD/NRMSE} & COV & $\langle\overline{|u|}\rangle$\\
\midrule
KS, periodic & SPOD-GPR & $\mathbf{8.14\times10^{-2}}$ & $\mathbf{1.26\times10^{-1}}$ & $0.160$ & {$0.65$} & $0.774$ & $2.76$\\
($\langle\overline{|u|}\rangle_{\mathrm{true}}=2.84$) & POD-GPR & $1.04\times10^{-1}$ & $1.45\times10^{-1}$ & $0.762$ & {$2.7$} & $0.989$ & $2.79$\\
\midrule
KS, travelling wave & SPOD-GPR & $\mathbf{0.238}$ & $\mathbf{0.295}$ & $0.520$ & {$0.90$} & $0.876$ & $1.81$\\
($\langle\overline{|u|}\rangle_{\mathrm{true}}=2.30$) & POD-GPR & $0.275$ & $0.327$ & $1.01$ & {$1.6$} & $0.926$ & $2.08$\\
\midrule
Cylinder, Reynolds & SPOD-GPR & $\mathbf{6.00\times10^{-3}}$ & $\mathbf{1.64\times10^{-2}}$ & $1.61\times10^{-2}$ & {$0.50$} & $0.723$ & $1.024$\\
($\langle\overline{|u|}\rangle_{\mathrm{true}}=1.026$) & POD-GPR & $2.54\times10^{-2}$ & $5.88\times10^{-2}$ & $6.62\times10^{-4}$ & {$5.7\times10^{-3}$} & $0.006$ & $0.999$\\
\midrule
Elliptic obstacle & SPOD-GPR & $\mathbf{2.48\times10^{-2}}$ & $\mathbf{5.81\times10^{-2}}$ & $0.327$ & {$2.9$} & $0.992$ & $1.000$\\
($\langle\overline{|u|}\rangle_{\mathrm{true}}=1.021$) & POD-GPR & $2.60\times10^{-2}$ & $6.05\times10^{-2}$ & $7.07\times10^{-3}$ & {$6.0\times10^{-2}$} & $0.043$ & $0.999$\\
\bottomrule
\end{tabular}
\end{table}

\subsection{Two-dimensional Kuramoto--Sivashinsky equation}
\label{sec:ks}

We study the anisotropic two-dimensional Kuramoto--Sivashinsky equation in the rescaled formulation of~\cite{kalogirou_-depth_2015},
\begin{equation}
u_t+\frac12\left(u_x^2+\alpha\,u_y^2\right)+u_{xx}+\alpha\,u_{yy}+\nu_1\left(\partial_x^2+\alpha\,\partial_y^2\right)^2u=0,\qquad \alpha=\frac{\nu_2}{\nu_1},
\label{eq:ks}
\end{equation}
posed on the periodic domain $(0,2\pi)^2$, with a zero spatial average and the initial condition $u_0=\sin(x+y)+\sin x+\sin y$. The parameters $\nu_1=(\pi/L_1)^2$ and $\nu_2=(\pi/L_2)^2$ indicate the dimensions $L_1\times L_2$ of the original domain; by varying these parameters one obtains steady, time-periodic, quasi-periodic or chaotic behaviour~\cite{kalogirou_-depth_2015}. The PDE is discretised using a $64\times64$ Fourier pseudo-spectral scheme and the time advancement is carried out with the exponential time-differencing fourth-order Runge--Kutta method (ETDRK4) using a time step of $\delta t_{\mathrm{FOM}}=0.1$. The weights $\mathbf{W}$ are taken to be the uniform quadrature weights corresponding to the periodic grid, and every fifth parameter setting is kept as test data. Two different regimes are studied, these being distinguished by the values of the parameters and by the resulting dynamical characteristics.
\subsubsection{Periodic regime}
\label{sec:ks-periodic}

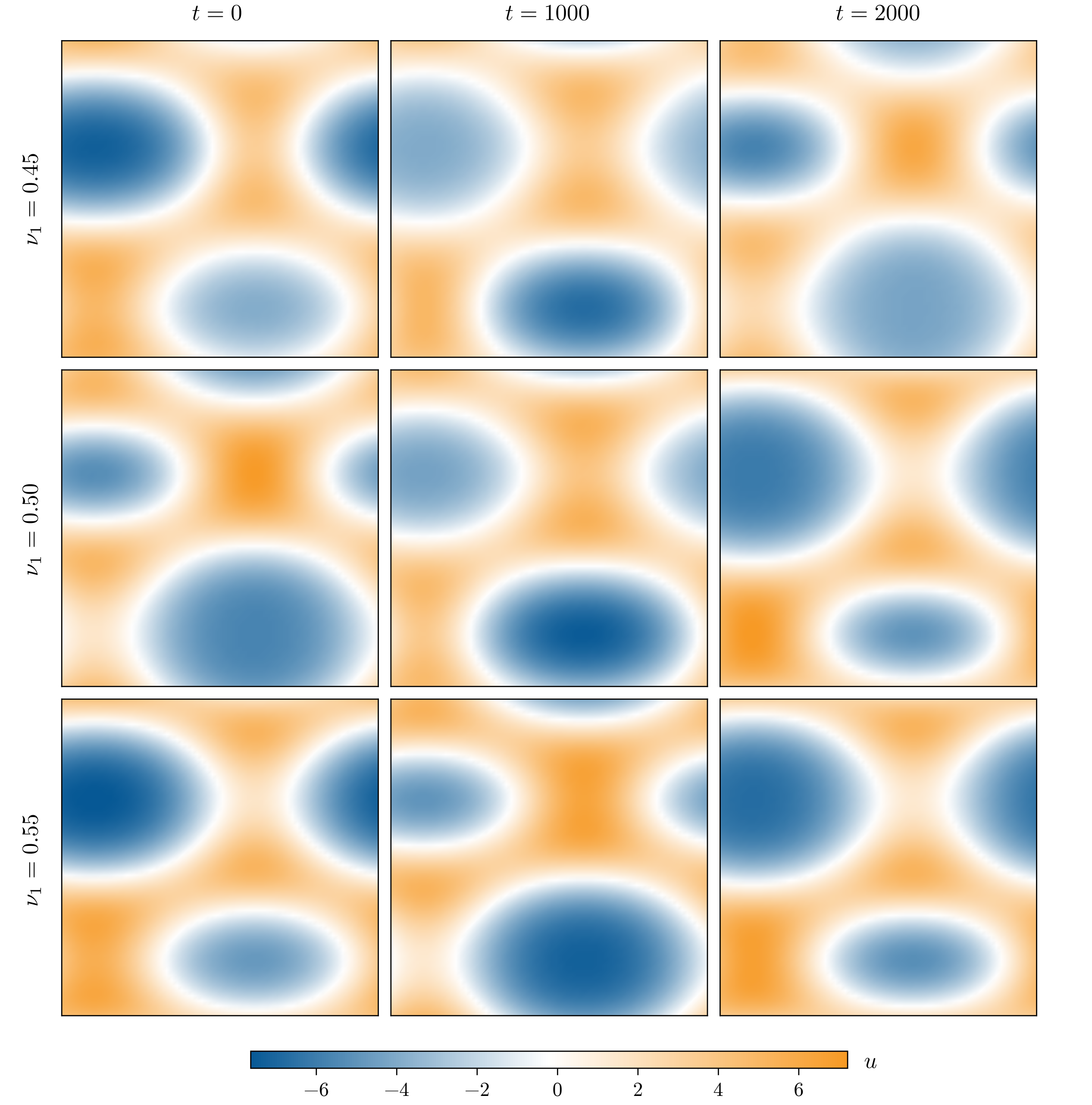
\begin{figure}[p]
\centering
\vspace*{1.2em}
\begin{overpic}[width=0.92\textwidth]{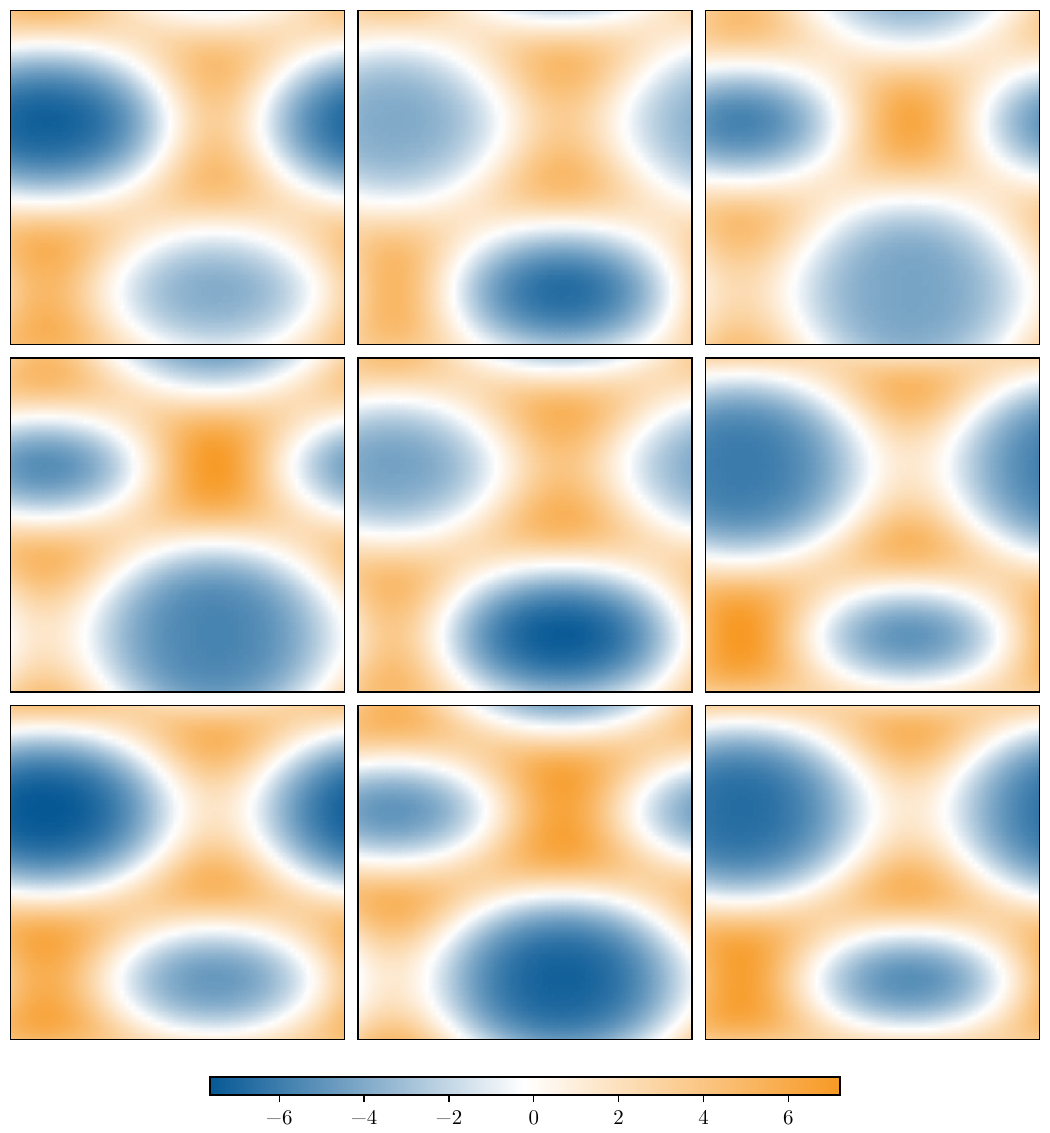}
\small
\put(15.4,101){\makebox(0,0)[b]{$t=0$}}
\put(46.2,101){\makebox(0,0)[b]{$t=1000$}}
\put(77.0,101){\makebox(0,0)[b]{$t=2000$}}
\put(-1,84.3){\rotatebox{90}{\makebox(0,0)[b]{$\nu_1=0.45$}}}
\put(-1,53.5){\rotatebox{90}{\makebox(0,0)[b]{$\nu_1=0.50$}}}
\put(-1,22.7){\rotatebox{90}{\makebox(0,0)[b]{$\nu_1=0.55$}}}
\put(75.7,3.8){\makebox(0,0)[l]{$u$}}
\end{overpic}
\caption{Kuramoto--Sivashinsky equation, periodic regime: full-order solution $u$ for three values of $\nu_1$ (rows) at three time instants (columns), with time measured from the first stored snapshot. }
\label{fig:ksp-maps-time}
\end{figure}

\begin{figure}[p]
\centering
\vspace*{1.2em}
\begin{overpic}[width=0.23\linewidth]{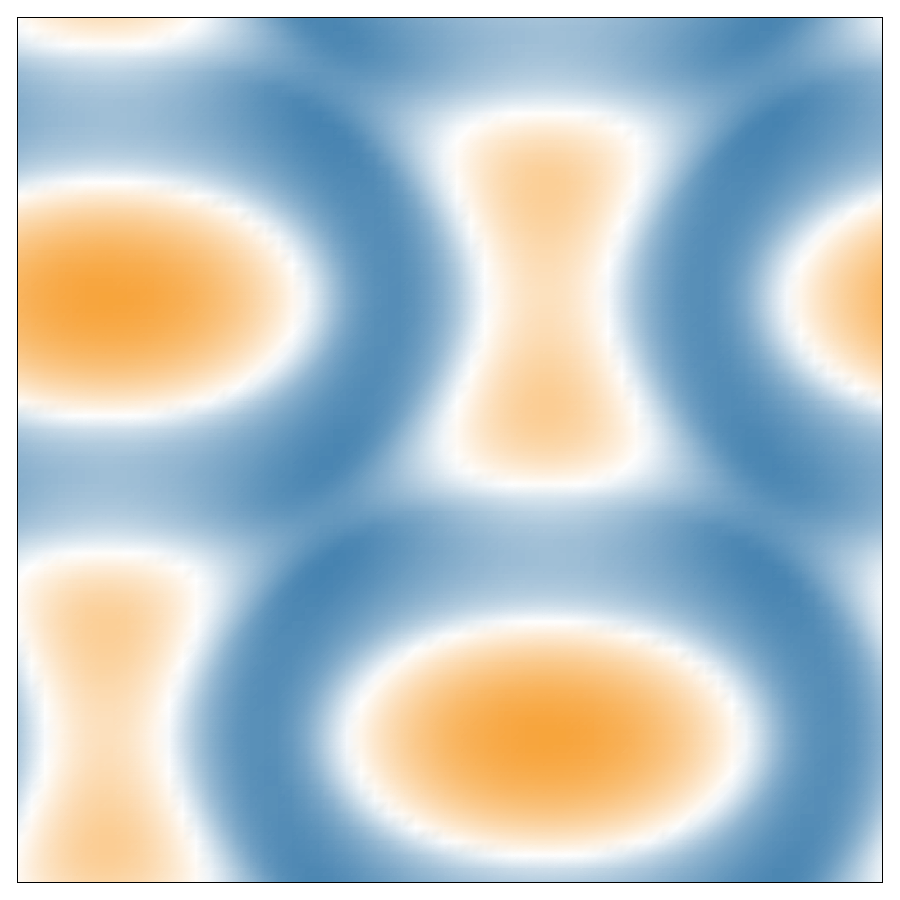}\small
\put(50,102){\makebox(0,0)[b]{Reference}}
\put(-4,50){\rotatebox{90}{\makebox(0,0)[b]{$\overline{|u|}$}}}
\end{overpic}\quad
\begin{overpic}[width=0.23\linewidth]{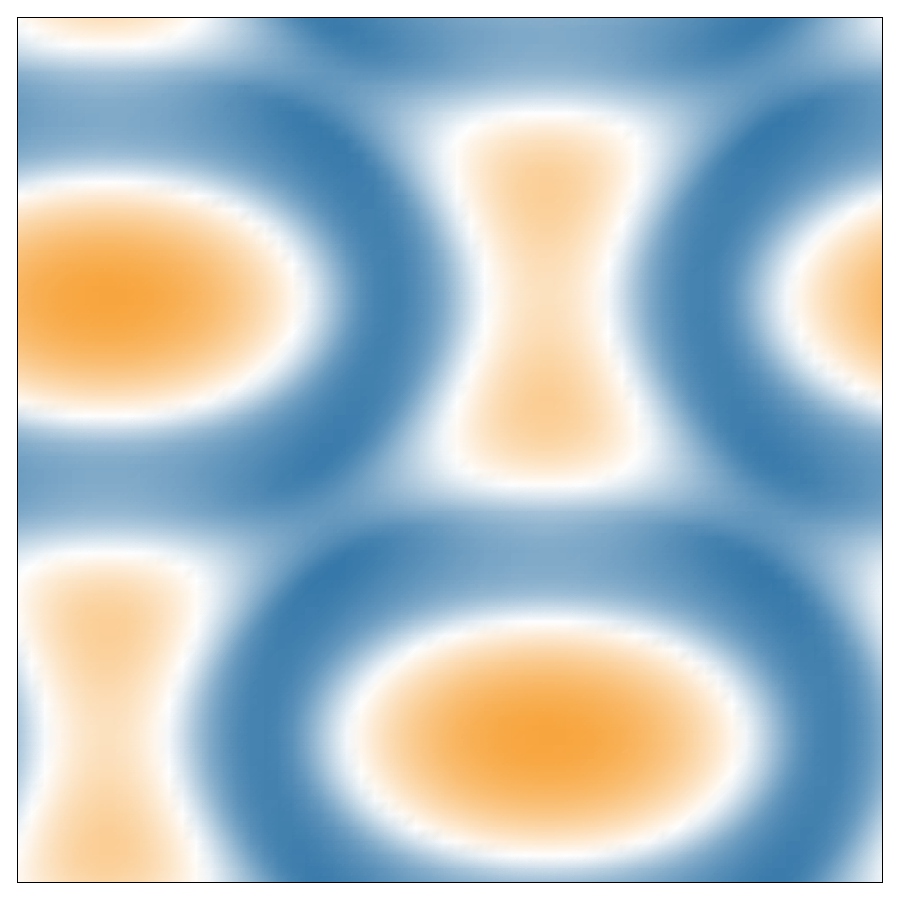}\small
\put(50,102){\makebox(0,0)[b]{SPOD-GPR}}
\end{overpic}\quad
\begin{overpic}[width=0.23\linewidth]{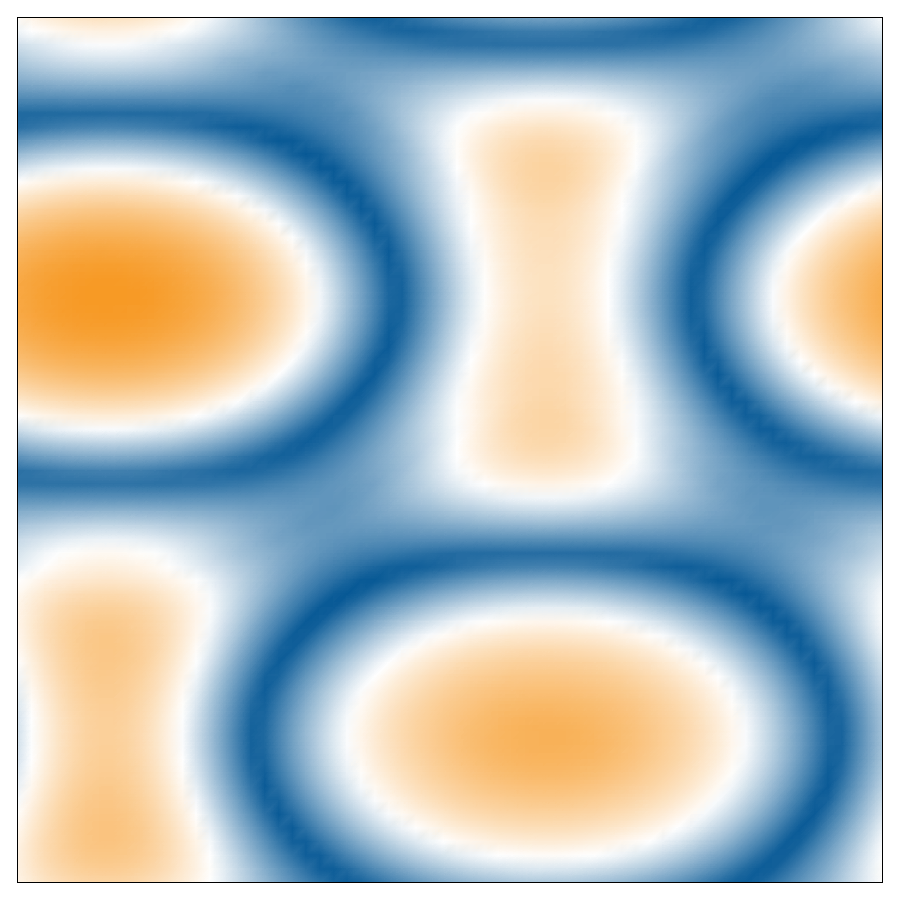}\small
\put(50,102){\makebox(0,0)[b]{POD-GPR}}
\end{overpic}\\[-2pt]
\includegraphics[width=0.5\linewidth]{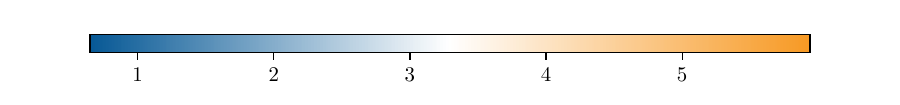}\\[16pt]
\begin{overpic}[width=0.23\linewidth]{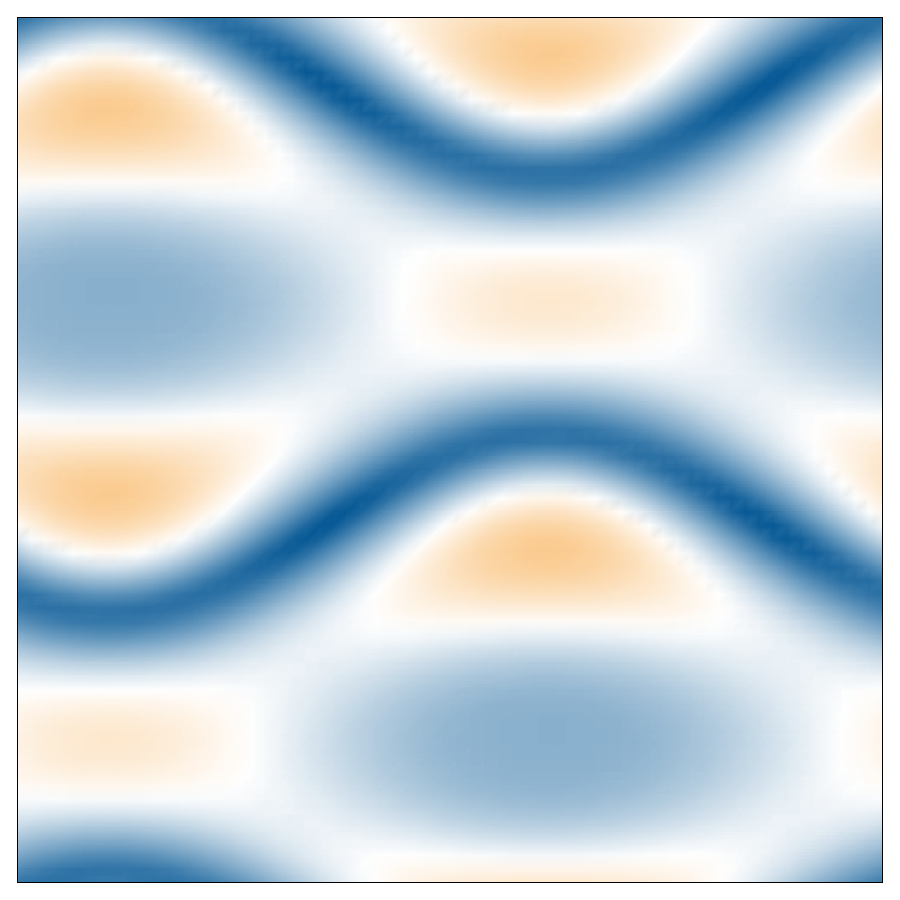}\small
\put(50,102){\makebox(0,0)[b]{SPOD-GPR}}
\put(-4,50){\rotatebox{90}{\makebox(0,0)[b]{NMAE}}}
\end{overpic}\qquad
\begin{overpic}[width=0.23\linewidth]{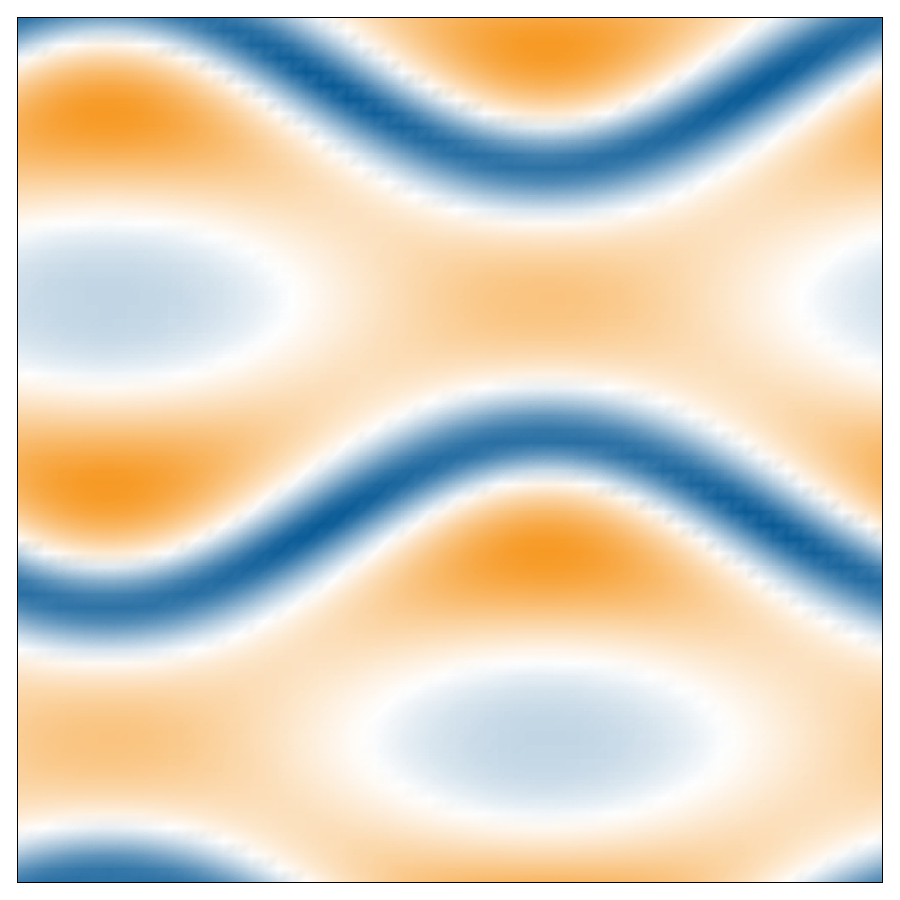}\small
\put(50,102){\makebox(0,0)[b]{POD-GPR}}
\end{overpic}\\[-2pt]
\includegraphics[width=0.5\linewidth]{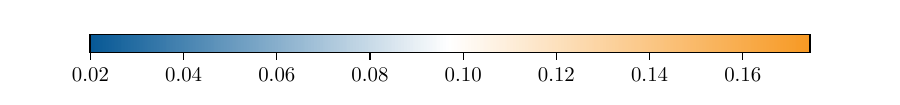}\\[4pt]
\begin{overpic}[width=0.23\linewidth]{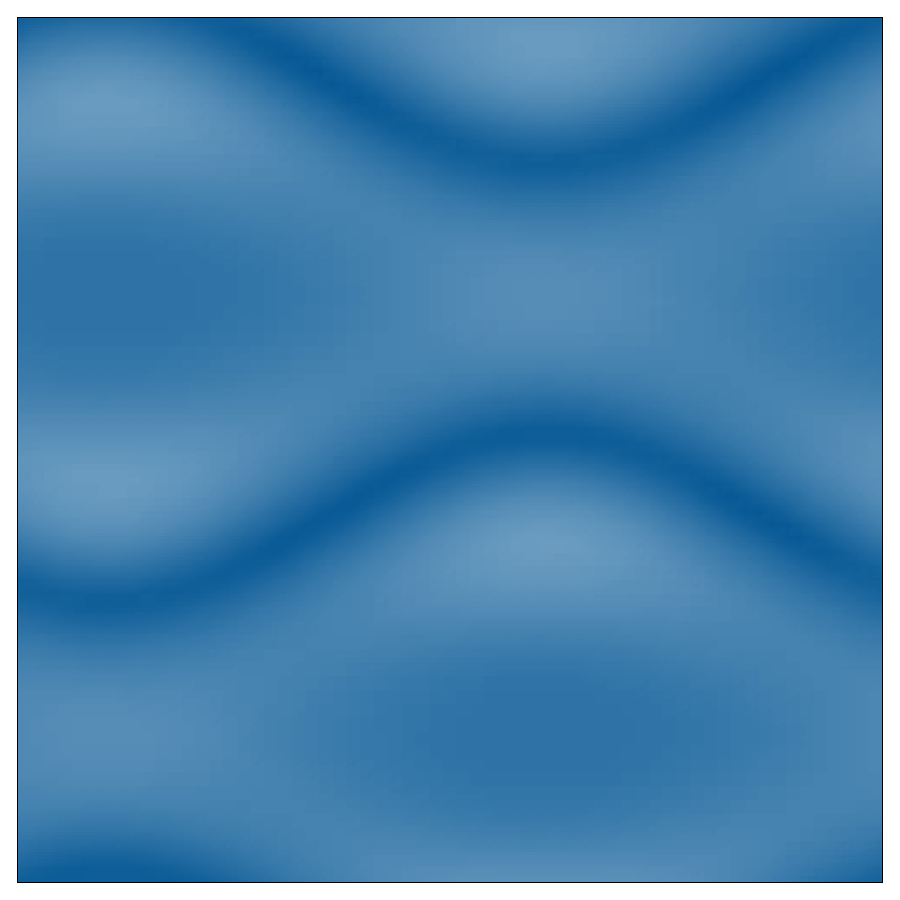}\small
\put(-4,50){\rotatebox{90}{\makebox(0,0)[b]{NSD}}}
\end{overpic}\qquad
\includegraphics[width=0.23\linewidth]{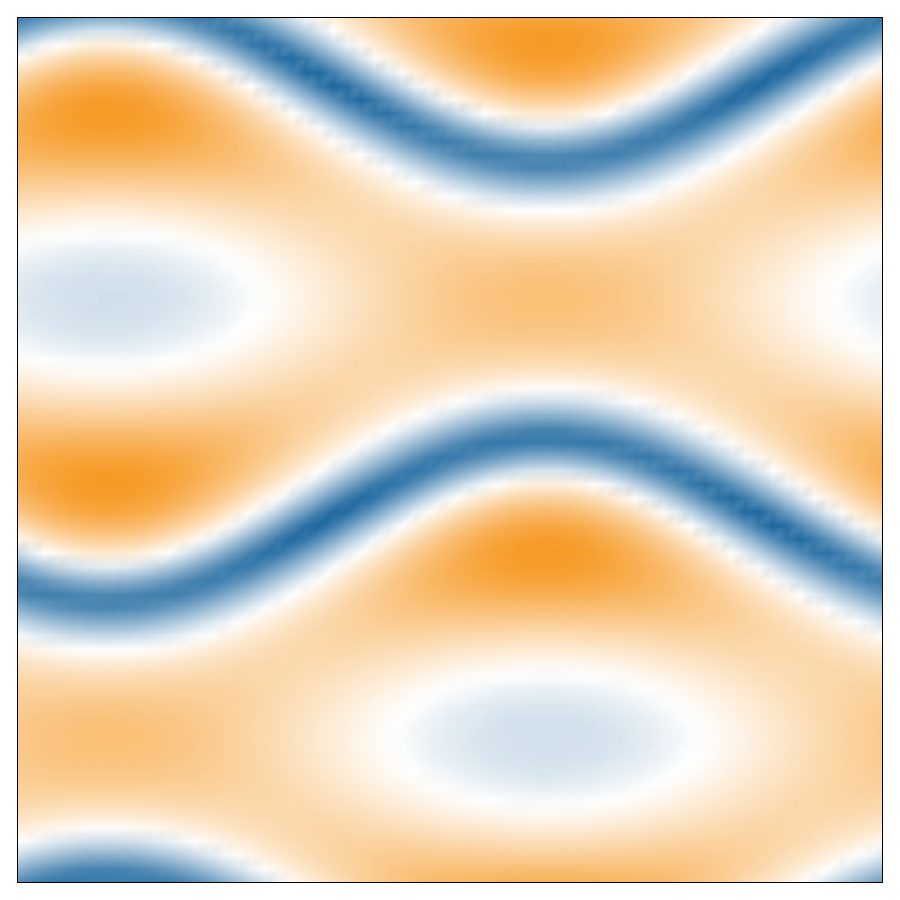}\\[-2pt]
\includegraphics[width=0.5\linewidth]{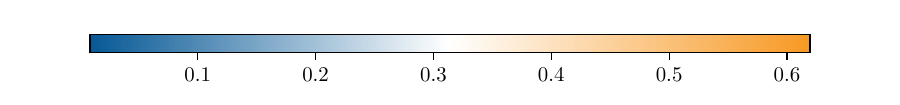}\\[4pt]
\begin{overpic}[width=0.23\linewidth]{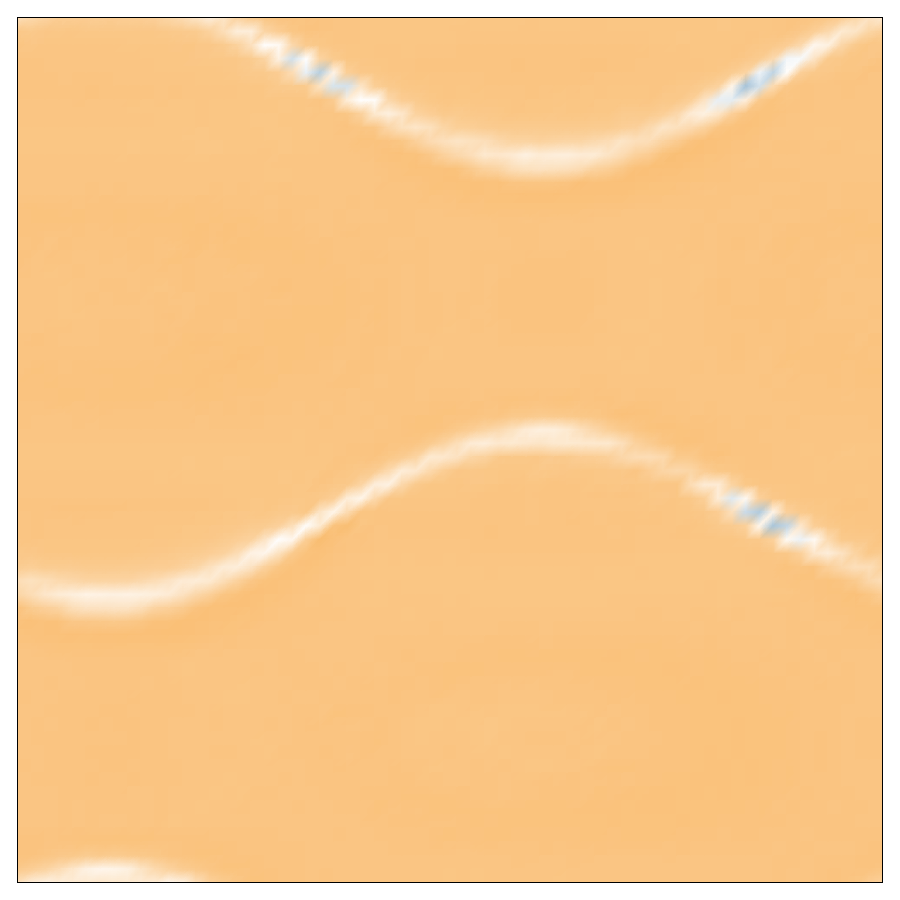}\small
\put(-4,50){\rotatebox{90}{\makebox(0,0)[b]{COV}}}
\end{overpic}\qquad
\includegraphics[width=0.23\linewidth]{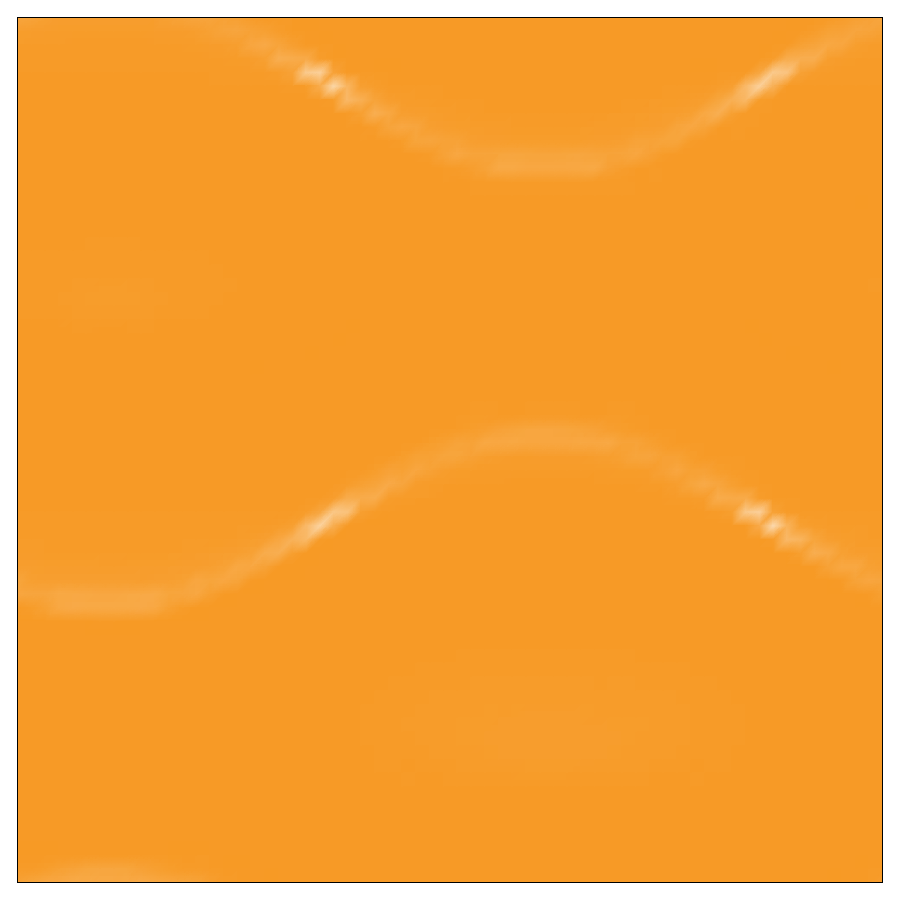}\\[-2pt]
\includegraphics[width=0.5\linewidth]{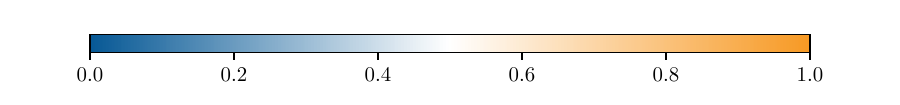}
\caption{Kuramoto--Sivashinsky equation, periodic regime ($\mu=\nu_1$): time-averaged maps covering the test window and test parameters{; NSD denotes the normalised predicted standard deviation and COV the empirical $95\%$ coverage}. The colour scale shown below each row is shared by the maps in that row.}
\label{fig:ksp-maps}
\end{figure}

In the first regime we take the value $\nu_2=0.2$ and select $\mu=\nu_1$ from the interval $[0.45,0.55]$. The dynamics that result show a spatially stationary cellular pattern with an amplitude which undergoes periodic temporal oscillations (as can be seen in \Cref{fig:ksp-maps-time} and \Cref{fig:ksp-maps}, top row). We integrate the equation until time $T=3000$, eliminate the first $t<1000$ transients, and store one snapshot every $20$ steps, thus obtaining $1000$ snapshots with a time step of $\delta t=2$; the test interval is $t\in[1900,1998]${, with time measured from the end of the discarded transient}. Since the time dependence is simple, SPOD-GPR needs only one mode for each frequency ($r=1$) and $K=3$ frequency bins, whereas POD-GPR uses $r_{\mathrm{POD}}=2$ POD modes.

\begin{figure}[t]
    \centering
    \includegraphics[width=\linewidth]{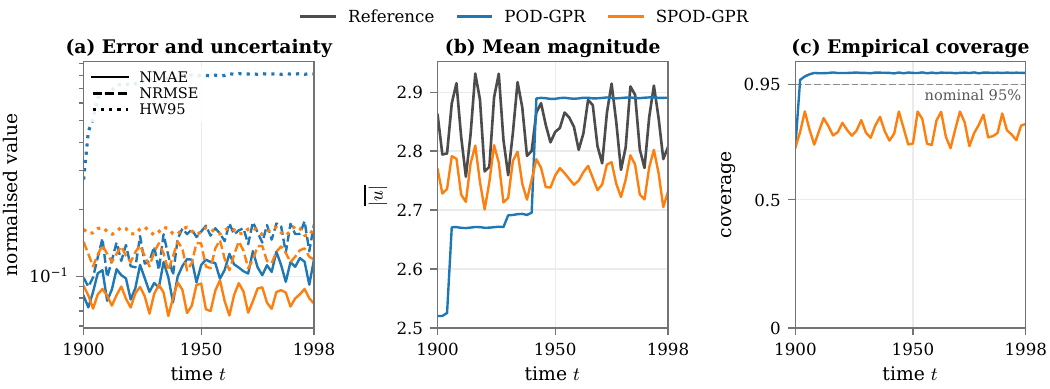}
    \caption{Kuramoto--Sivashinsky equation in the periodic regime ($\mu=\nu_1$), test window. On the left: the NMAE (solid), the NRMSE (dashed) and the predicted normalised $95\%$ half-width (dotted) for SPOD-GPR ({orange}) and POD-GPR ({blue}){; for a calibrated Gaussian predictive distribution, the half-width is about $1.96$ times the NRMSE}. In the centre: the space average of the mean magnitude $\overline{|u|}${, with the reference solution in black}. On the right: the empirical coverage of the $95\%$ predictive interval.}
    \label{fig:ksp-time}
\end{figure}

The time traces shown in \Cref{fig:ksp-time} show the main qualitative difference between the two surrogate models. The space-averaged magnitude from SPOD-GPR stays phase-aligned with the true values for the whole of the prediction period, whereas the output from POD-GPR quickly settles down after just a few steps to a level that is essentially constant\ {, apart from a few step-like jumps between plateaus}: it seems that the autoregressive propagation of the POD coefficients is approaching a fixed point and thus the oscillating behaviour is eliminated. Both methods slightly underpredict the amplitude (the space--time means are $2.76$ and $2.79$ as compared to $2.84$). As stated in \Cref{tab:results}, SPOD-GPR reduces the NMAE by about $22\%$ and the NRMSE by around $13\%$; its error varies periodically with the solution. The spatial fields in \Cref{fig:ksp-maps} show that the errors of both models reflect the cellular structure, with that of SPOD-GPR being smaller almost everywhere.

In this case, the uncertainty quantification is not very convincing. The prediction interval from the SPOD-GPR method is too narrow (the normalised half-width is $0.16$ as against an NRMSE of $0.13${, i.e.\ NSD/NRMSE${}\approx0.65$}), the coverage therefore being $0.77$; it tends towards one mainly along the nodal lines of the pattern, where the error is small, and is lower in the other areas. POD-GPR shows the opposite behaviour: its interval is very wide (half-width $0.76$, about five times its NRMSE{, i.e.\ NSD/NRMSE${}\approx2.7$}), so that its coverage of $0.99$ is obtained at the price of a weakly informative band. Although the spatial distribution of the POD-GPR standard deviation follows that of its error, its amplitude is overestimated.

\subsubsection{Travelling-wave regime}
\label{sec:ks-wave}

\begin{figure}[p]
\centering
\vspace*{1.2em}
\begin{overpic}[width=0.92\textwidth]{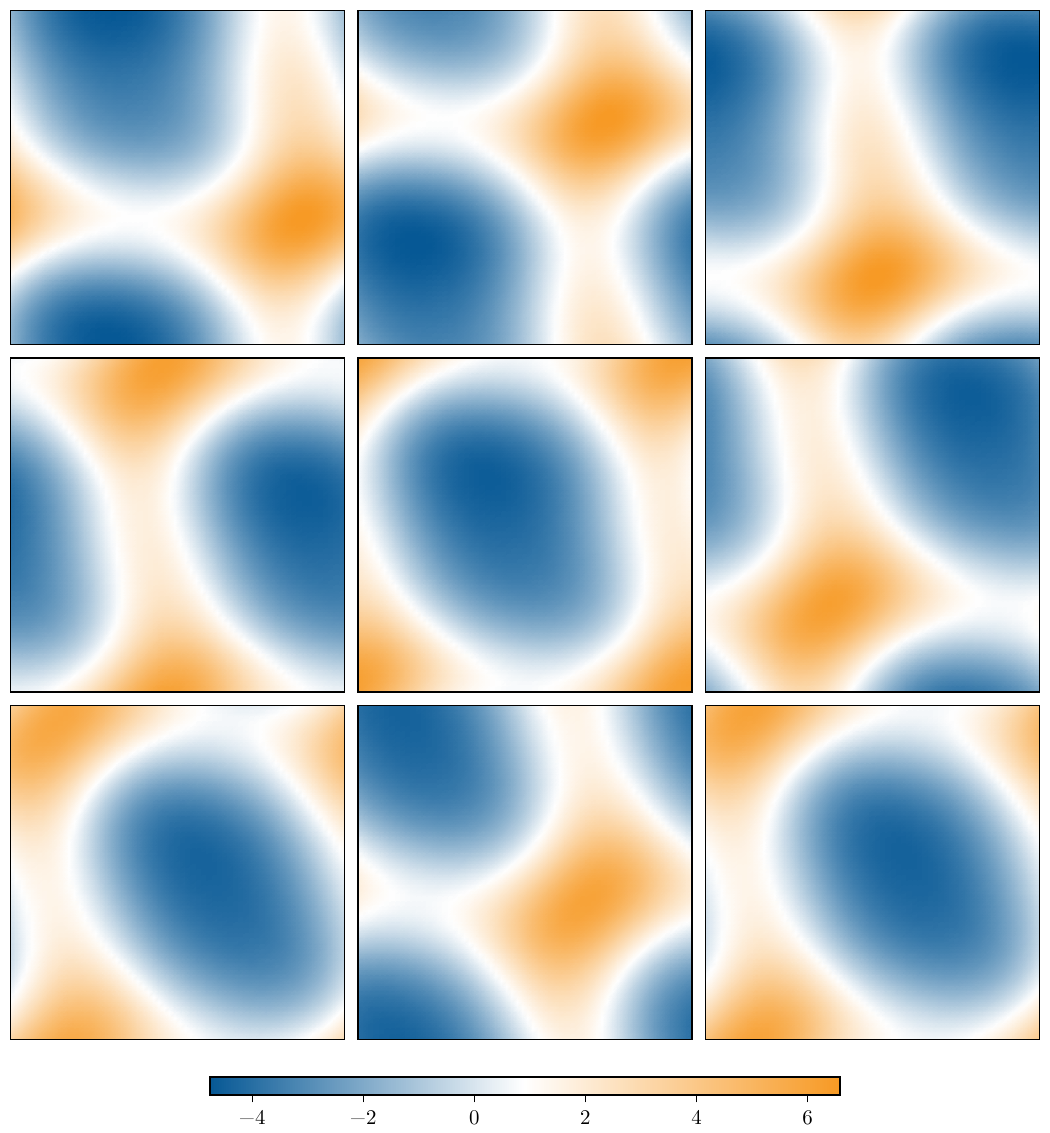}
\small
\put(15.4,101){\makebox(0,0)[b]{$t=0$}}
\put(46.2,101){\makebox(0,0)[b]{$t=2000$}}
\put(77.0,101){\makebox(0,0)[b]{$t=4000$}}
\put(-1,84.3){\rotatebox{90}{\makebox(0,0)[b]{$\nu_2=0.30$}}}
\put(-1,53.5){\rotatebox{90}{\makebox(0,0)[b]{$\nu_2=0.35$}}}
\put(-1,22.7){\rotatebox{90}{\makebox(0,0)[b]{$\nu_2=0.40$}}}
\put(75.7,3.8){\makebox(0,0)[l]{$u$}}
\end{overpic}
\caption{Kuramoto--Sivashinsky equation, travelling-wave regime{: full-order solution $u$ for three values of $\nu_2$ (rows) at three time instants (columns), with time measured from the first stored snapshot}. }
\label{fig:ks-maps-time}
\end{figure}

In the second regime we take the value $\nu_1=0.5$ and select $\mu=\nu_2$ from the interval $[0.3,0.4]$. For values in this range the solution forms quasi-periodic, travelling-wave-type coherent structures having a more complicated temporal spectrum. We integrate the equation until $T=4100$; after having eliminated the initial transient for $t<100.1$ we take one snapshot every $40$ steps, which gives a total of $1000$ snapshots with a time step $\delta t=4$. The test time interval is $t\in[3900.1,4096.1]${, in absolute time}. We use $K=20$ frequency bins and retain $r=4$ modes in each bin, while the POD-GPR is constructed from $r_{\mathrm{POD}}=6$ POD modes.
The configurations at different times and parameters are shown in \Cref{fig:ks-maps-time}.

\begin{figure}[t]
    \centering
    \includegraphics[width=\linewidth]{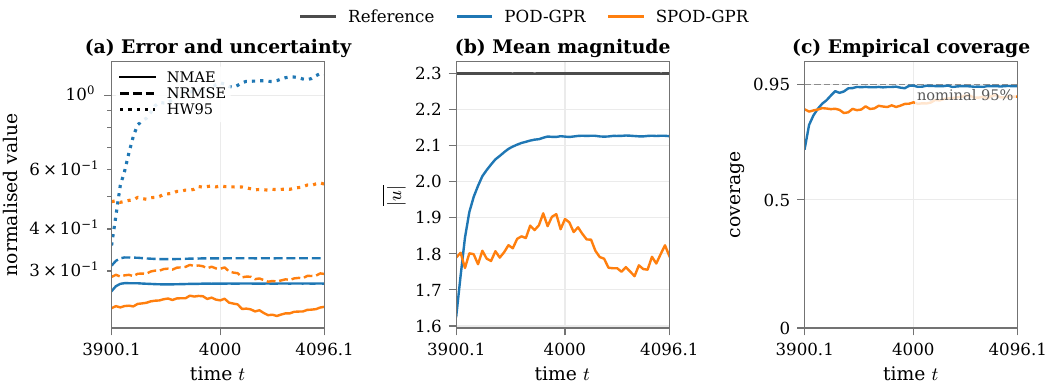}
    \caption{Kuramoto--Sivashinsky equation in the travelling-wave regime ($\mu=\nu_2$), test window. On the left: the NMAE (solid), the NRMSE (dashed) and the predicted normalised $95\%$ half-width (dotted) for SPOD-GPR ({orange}) and POD-GPR ({blue}){; for a calibrated Gaussian predictive distribution, the half-width is about $1.96$ times the NRMSE}. In the centre: the space average of the mean magnitude $\overline{|u|}${, with the reference solution in black}. On the right: the empirical coverage of the $95\%$ predictive interval.}
    \label{fig:ks-time}
\end{figure}

\begin{figure}[p]
\centering
\vspace*{1.2em}
\begin{overpic}[width=0.23\linewidth]{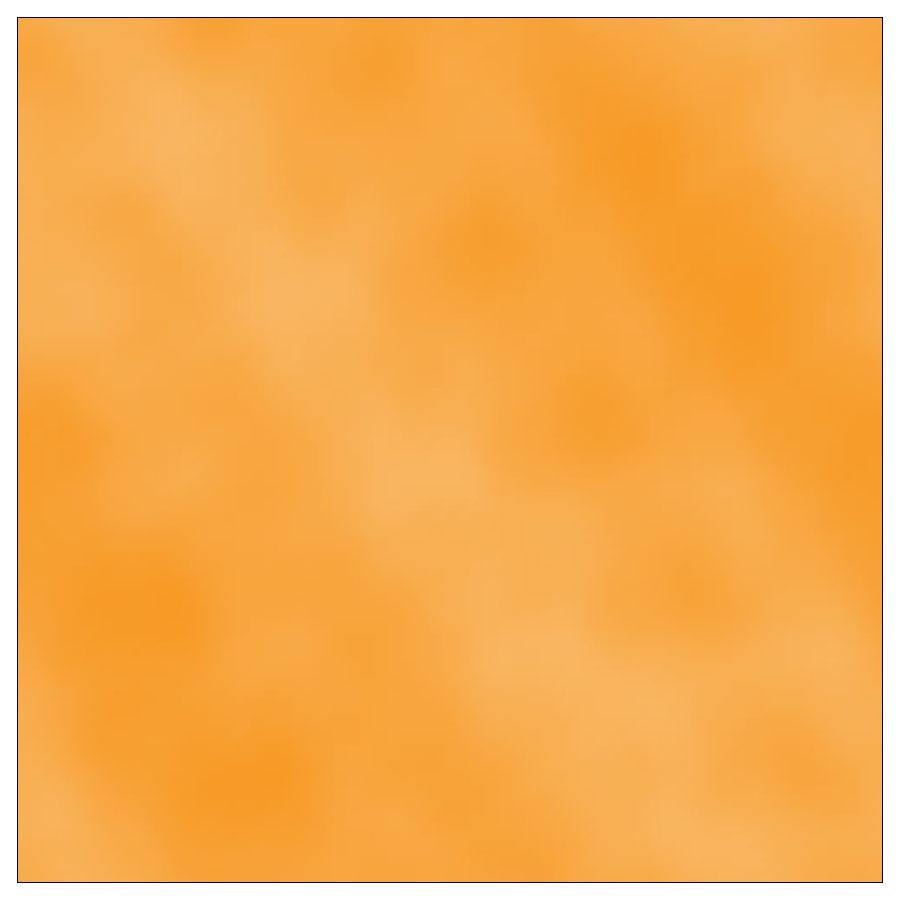}\small
\put(50,102){\makebox(0,0)[b]{Reference}}
\put(-4,50){\rotatebox{90}{\makebox(0,0)[b]{$\overline{|u|}$}}}
\end{overpic}\quad
\begin{overpic}[width=0.23\linewidth]{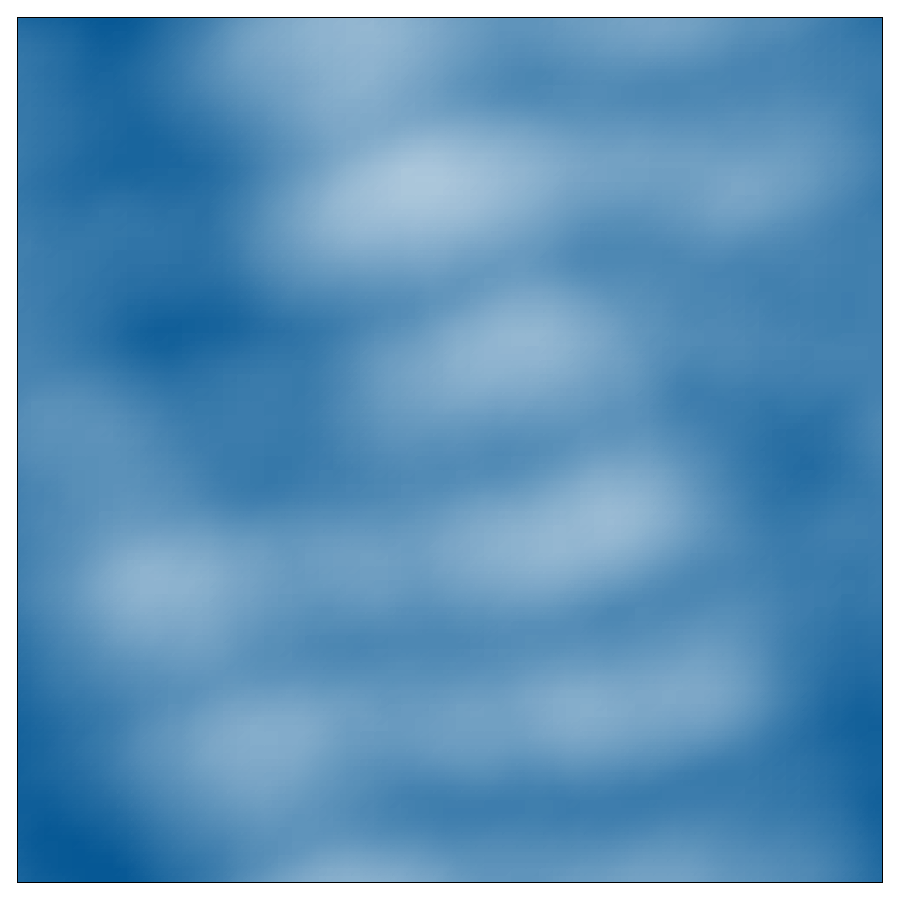}\small
\put(50,102){\makebox(0,0)[b]{SPOD-GPR}}
\end{overpic}\quad
\begin{overpic}[width=0.23\linewidth]{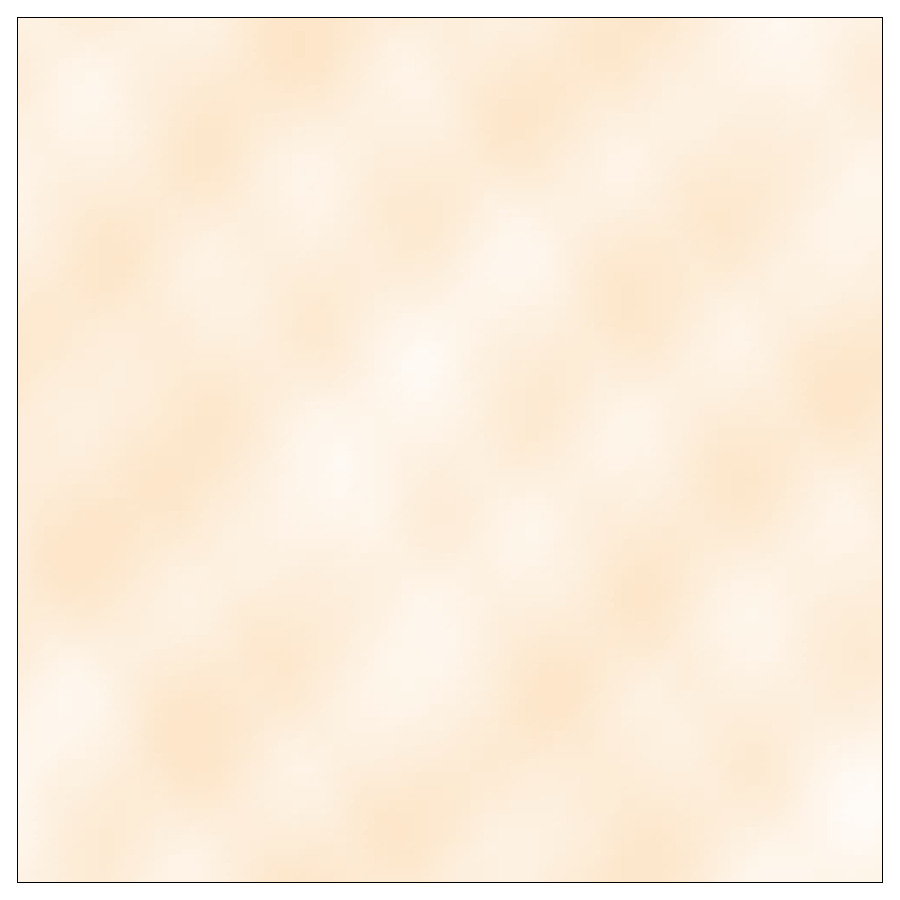}\small
\put(50,102){\makebox(0,0)[b]{POD-GPR}}
\end{overpic}\\[-2pt]
\includegraphics[width=0.5\linewidth]{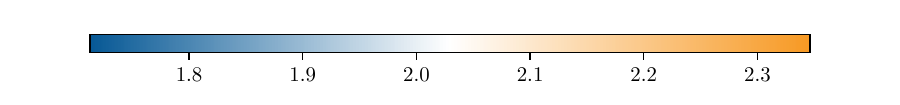}\\[16pt]
\begin{overpic}[width=0.23\linewidth]{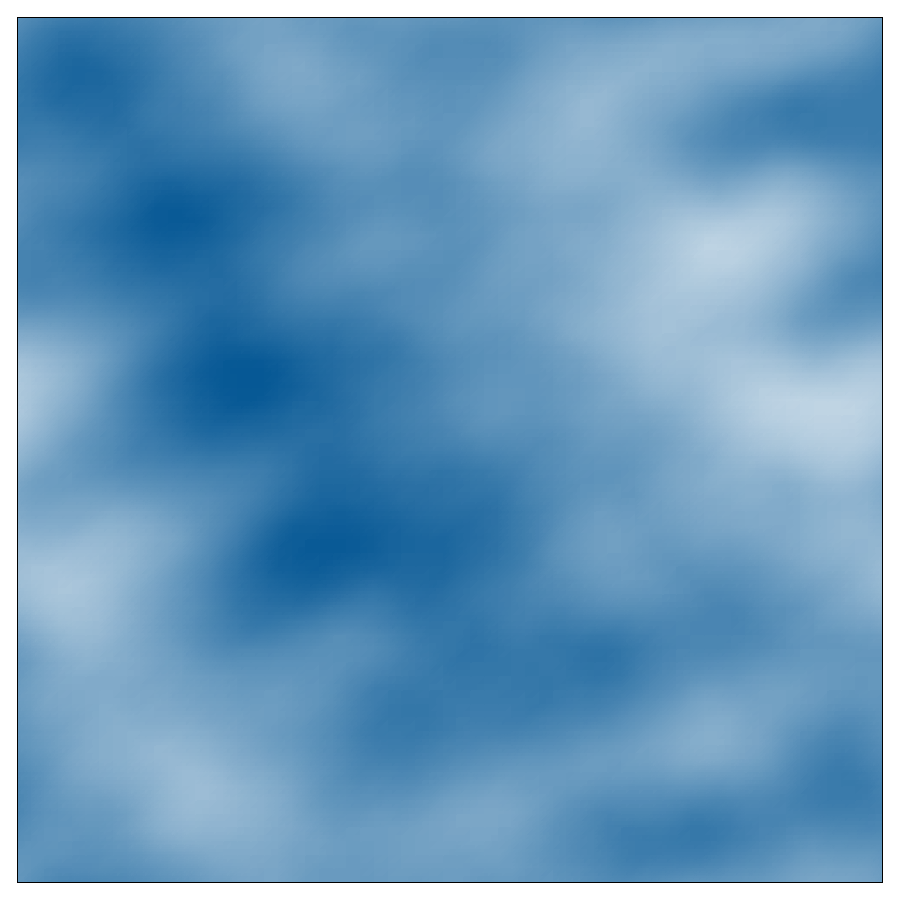}\small
\put(50,102){\makebox(0,0)[b]{SPOD-GPR}}
\put(-4,50){\rotatebox{90}{\makebox(0,0)[b]{NMAE}}}
\end{overpic}\qquad
\begin{overpic}[width=0.23\linewidth]{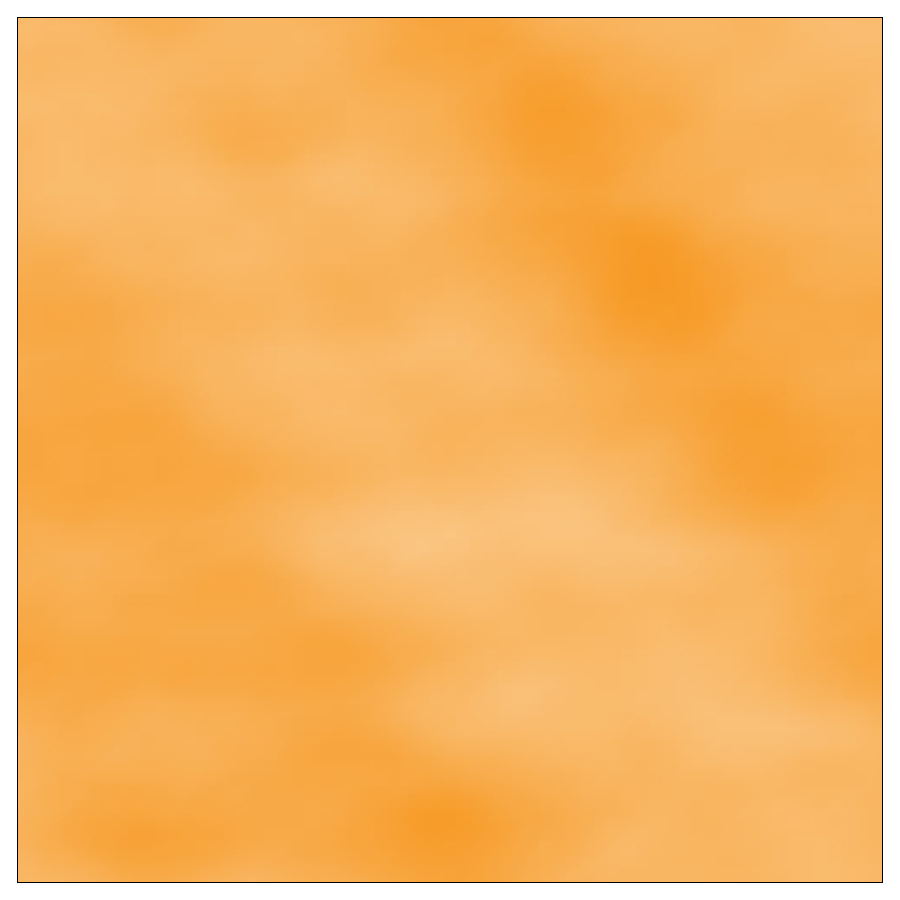}\small
\put(50,102){\makebox(0,0)[b]{POD-GPR}}
\end{overpic}\\[-2pt]
\includegraphics[width=0.5\linewidth]{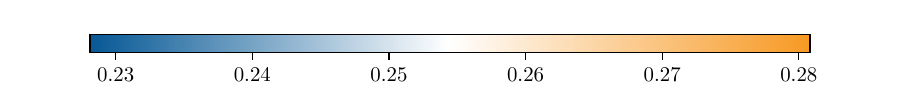}\\[4pt]
\begin{overpic}[width=0.23\linewidth]{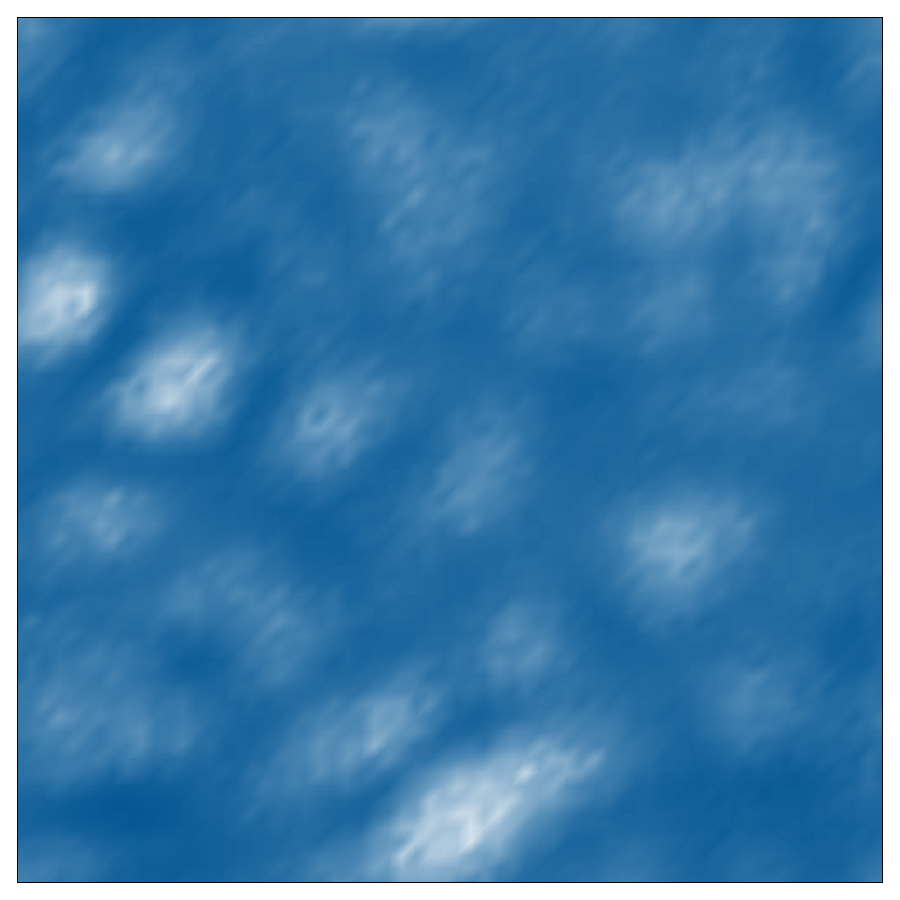}\small
\put(-4,50){\rotatebox{90}{\makebox(0,0)[b]{NSD}}}
\end{overpic}\qquad
\includegraphics[width=0.23\linewidth]{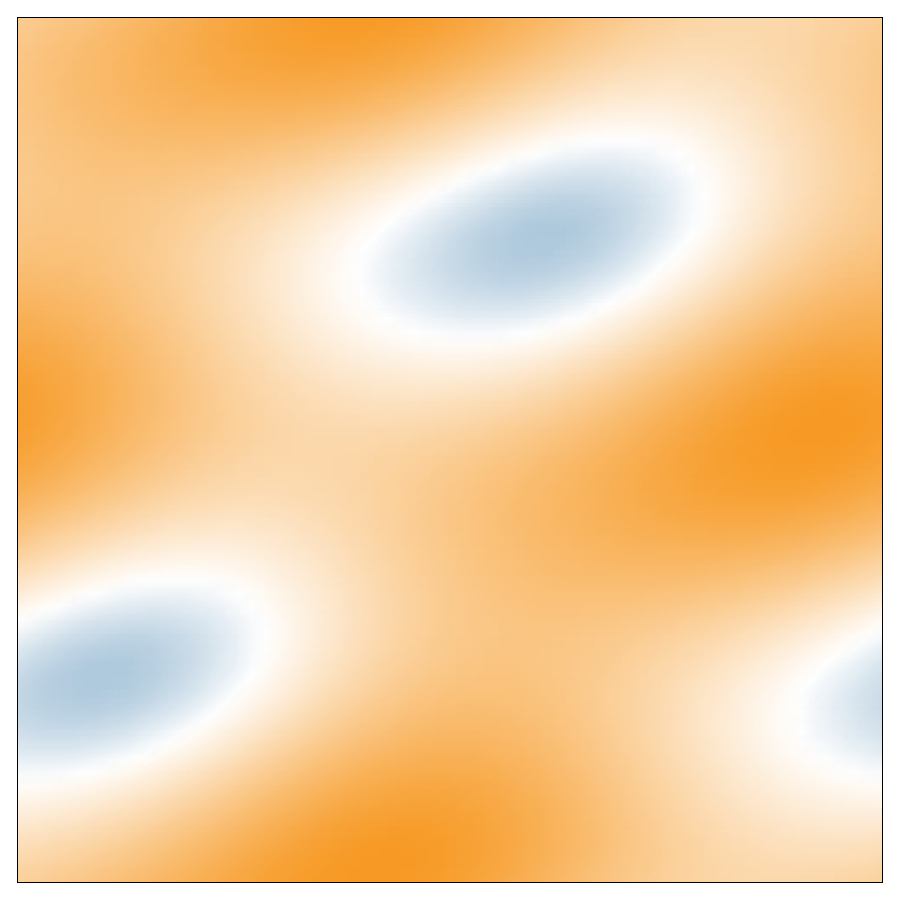}\\[-2pt]
\includegraphics[width=0.5\linewidth]{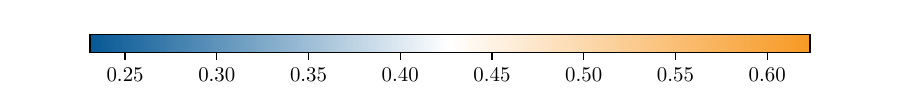}\\[4pt]
\begin{overpic}[width=0.23\linewidth]{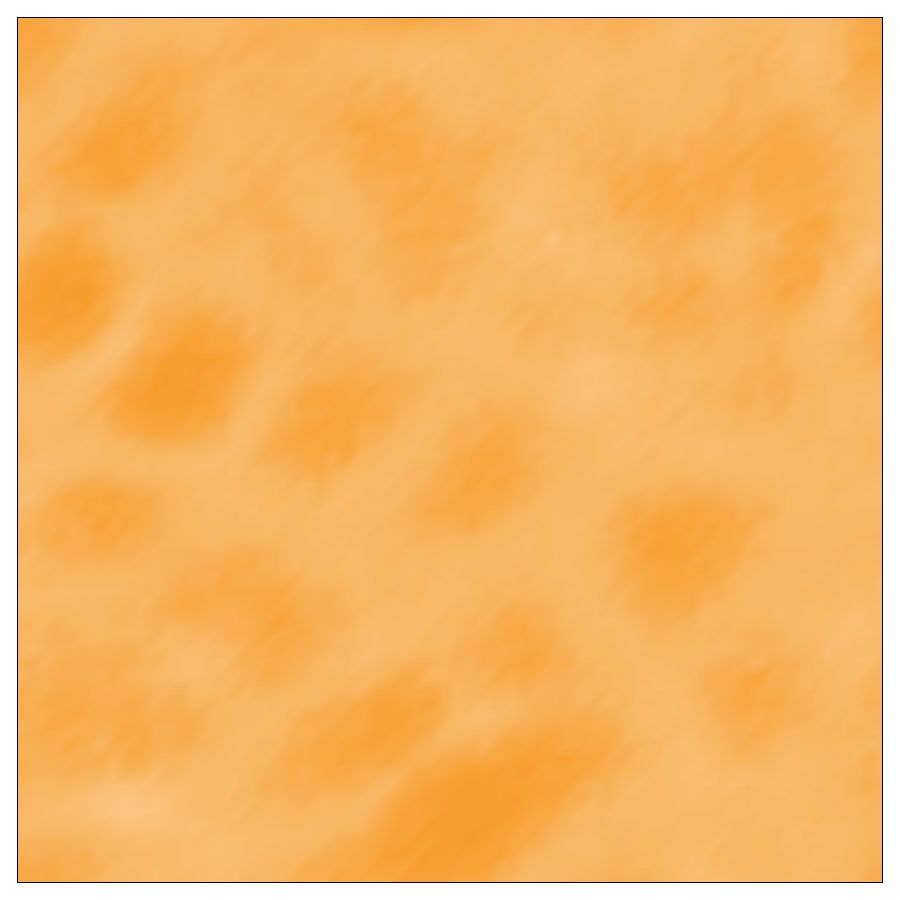}\small
\put(-4,50){\rotatebox{90}{\makebox(0,0)[b]{COV}}}
\end{overpic}\qquad
\includegraphics[width=0.23\linewidth]{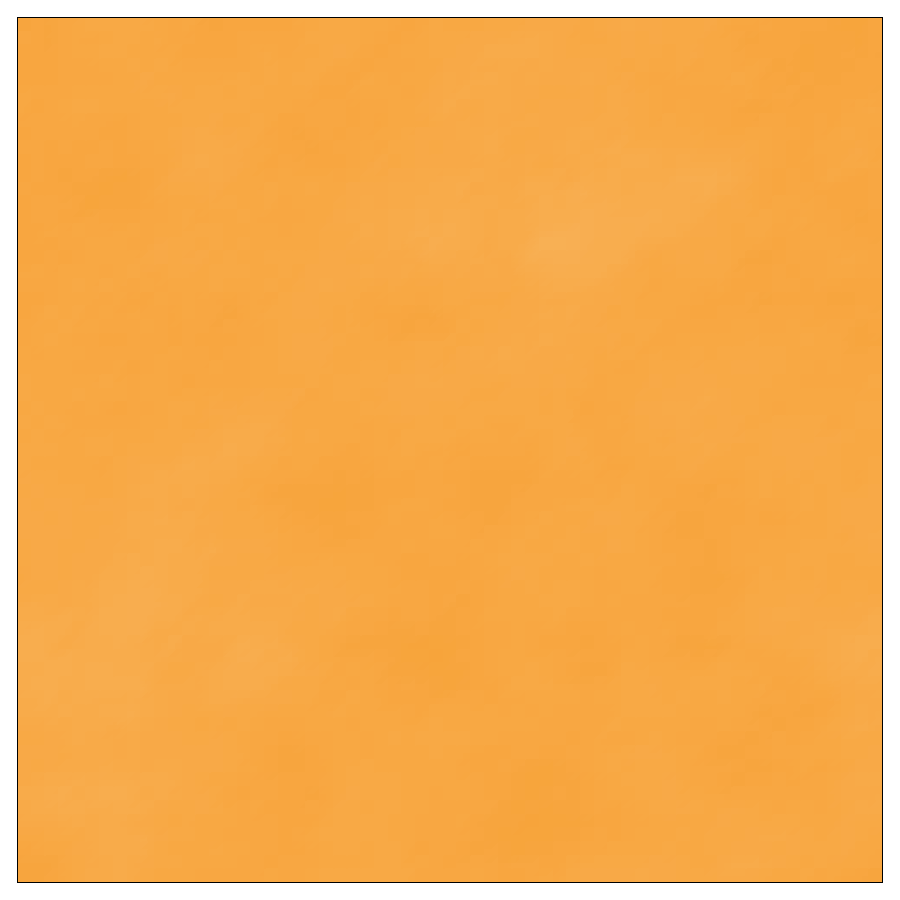}\\[-2pt]
\includegraphics[width=0.5\linewidth]{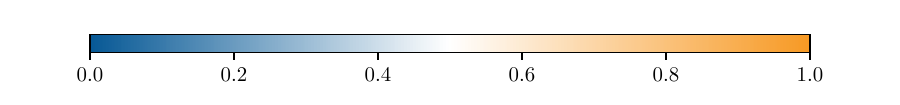}
\caption{Kuramoto--Sivashinsky equation, travelling-wave regime ($\mu=\nu_2$): time-averaged maps covering the test window and test parameters{; NSD denotes the normalised predicted standard deviation and COV the empirical $95\%$ coverage}. The colour scale shown below each row is shared by the maps in that row{; note the narrow range of the NMAE colour scale}.}
\label{fig:ks-maps}
\end{figure}

\Cref{fig:ks-time} illustrates how the errors, the predicted uncertainty, the coverage, and the mean magnitude develop over time. Throughout the entire test period, SPOD-GPR has a smaller NMAE and NRMSE than POD-GPR—by about $13\%$ and $10\%$ respectively (see \Cref{tab:results}). Both methods underestimate the field amplitude, that is, their posterior means are too damped, which is to be expected when dealing with {quasi-periodic travelling waves predicted at unseen parameters and beyond the training window}: once phase coherence is lost, the prediction which minimises the error usually moves back towards the mean. Interestingly, although SPOD-GPR predicts an even smaller amplitude than POD-GPR it still manages to achieve a lower pointwise error, {which suggests a better phase preservation, although pointwise error measures also tend to favour more damped predictions once the phase is lost}. The uncertainty from POD-GPR increases rapidly and reaches a $95\%$ half-width comparable to the full data range, consistent with its high coverage ($0.93$), while SPOD-GPR predicts an interval that is roughly half as wide and has a slightly lower coverage ($0.88$){; in terms of the spread--skill ratio, SPOD-GPR is close to calibration (NSD/NRMSE${}\approx0.90$), whereas POD-GPR is conservative (${}\approx1.6$)}. The {time-averaged maps} in \Cref{fig:ks-maps} confirm that SPOD-GPR gives consistently smaller errors and that its predicted standard deviation follows the structure of the travelling patterns, rather than the smooth, weakly informative envelope produced by POD-GPR.

\subsection{Flow past a cylinder: Reynolds-number parametrisation}
\label{sec:cyl}

A second benchmark involves incompressible flow past a circular cylinder of diameter $D=1$ which is situated at $(5,5)$ in the channel $(0,40)\times(0,10)$. The flow is given by the Navier--Stokes system
\begin{equation}
\partial_t\mathbf{u}+(\mathbf{u}\cdot\nabla)\mathbf{u}-\frac{1}{Re}\Delta\mathbf{u}+\nabla p=\mathbf{0},\qquad \nabla\cdot\mathbf{u}=0,
\end{equation}
together with a parabolic inflow of unit mean speed, $u_x=6y(H-y)/H^2$ with $H=10$, no-slip boundary conditions on both the channel walls and the cylinder surface, and zero pressure at the outlet. The Reynolds number $Re\in[110,160]$, defined in terms of the mean inflow velocity and the cylinder diameter, is taken as the parameter. The Reynolds numbers are selected uniformly across this range, the total number of samples being $100$, $80$ of which are used for training and $20$ for testing.

\begin{figure}[tbp]
\centering
\vspace*{1.2em}
\begin{overpic}[width=0.82\textwidth]{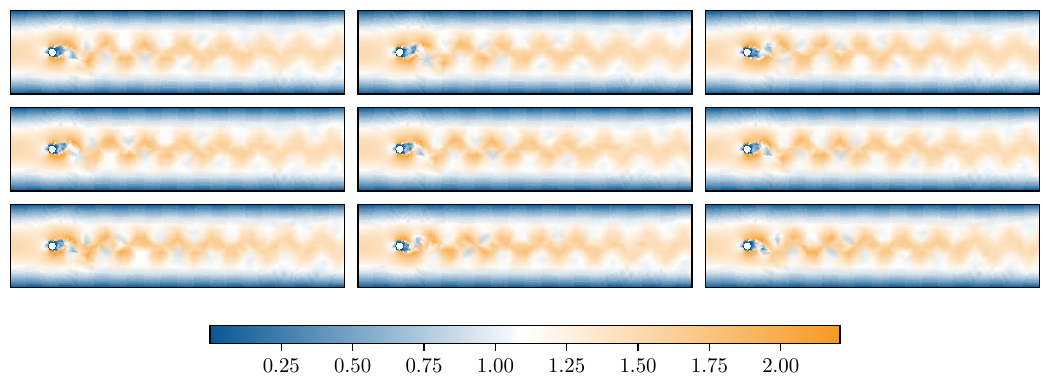}
\small
\put(16.8,37){\makebox(0,0)[b]{$t=0$}}
\put(49.85,37){\makebox(0,0)[b]{$t=50$}}
\put(82.95,37){\makebox(0,0)[b]{$t=100$}}
\put(-0.5,31.15){\makebox(0,0)[r]{$Re=110$}}
\put(-0.5,21.95){\makebox(0,0)[r]{$Re=135$}}
\put(-0.5,12.75){\makebox(0,0)[r]{$Re=160$}}
\put(81,4.25){\makebox(0,0)[l]{$|\mathbf{u}|$}}
\end{overpic}
\caption{Flow past a cylinder, Reynolds-number parametrisation{: full-order velocity magnitude $|\mathbf{u}|$ for three values of $Re$ (rows) at three time instants (columns), with time measured from the first stored snapshot}.}
\label{fig:cyl-maps-time}
\end{figure}

The PDEs are advanced by means of an incremental pressure-correction method using Adams--Bashforth/Crank--Nicolson time stepping and $\mathbb{Q}_2$--$\mathbb{Q}_1$ finite elements on a curved quadrilateral mesh which is refined in the wake region, the implementation being carried out in DOLFINx~\cite{baratta_dolfinx_2023}, with a time step $\delta t=0.1$. After $1000$ initial steps have been taken to enable the vortex street to form, an additional $1000$ steps ($t\in(100,200]$) are stored and then interpolated in space to a coarser, quasi-uniform $\mathbb{Q}_1$ mesh that is used by the ROMs. The weight matrix $\mathbf{W}$ is taken as the lumped $\mathbb{Q}_1$ mass matrix on this mesh, and this matrix is duplicated for each velocity component. The testing interval $t\in(195,200]$ corresponds to approximately one vortex-shedding cycle, while the SPOD window $N_f\,\delta t=10.1$ is equivalent to about two cycles; only the $K=3$ highest-energy frequency bins—given the window resolution ($1/(N_f\delta t)\approx0.1$), these are concentrated near the shedding frequency—and $r=2$ modes per bin are retained, while POD-GPR uses $r_{\mathrm{POD}}=2$ POD modes. {Representative snapshots of the velocity magnitude are shown in \Cref{fig:cyl-maps-time}.}

\begin{figure}[t]
    \centering
    \includegraphics[width=\linewidth]{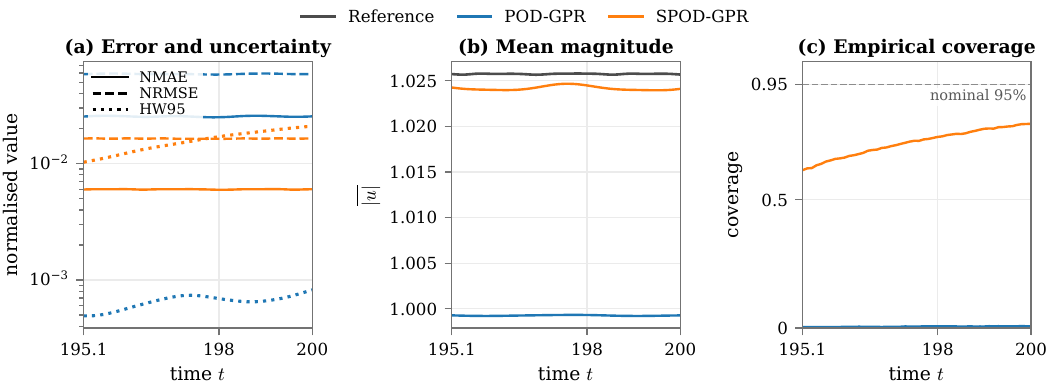}
    \caption{Flow past a cylinder, Reynolds-number parametrisation, test window. {On the left: the NMAE (solid), the NRMSE (dashed) and the predicted normalised $95\%$ half-width (dotted) for SPOD-GPR (orange) and POD-GPR (blue); for a calibrated Gaussian predictive distribution, the half-width is about $1.96$ times the NRMSE. In the centre: the space average of the mean velocity magnitude $\overline{|\mathbf{u}|}$, with the reference solution in black. On the right: the empirical coverage of the $95\%$ predictive interval.}}
    \label{fig:cyl-time}
\end{figure}

\begin{figure}[p]
\centering
\vspace*{1.2em}
\begin{overpic}[width=0.46\linewidth]{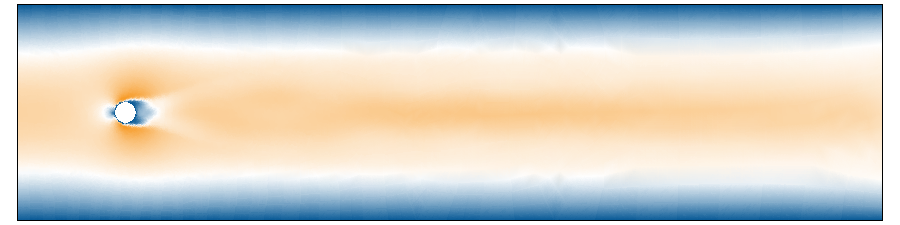}\small
\put(50,26.5){\makebox(0,0)[b]{Reference}}
\put(-2,12.5){\rotatebox{90}{\makebox(0,0)[b]{$\overline{|\mathbf{u}|}$}}}
\end{overpic}\\[14pt]
\hspace*{1.3em}\begin{overpic}[width=0.46\linewidth]{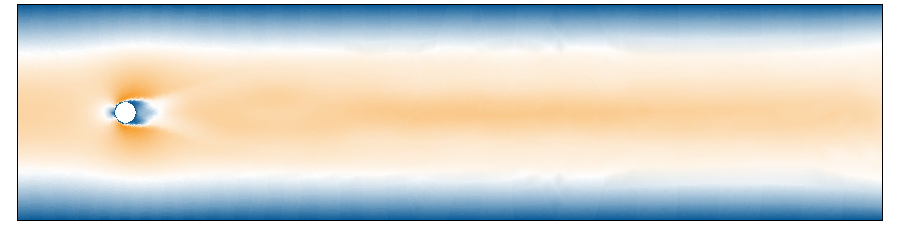}\small
\put(50,26.5){\makebox(0,0)[b]{SPOD-GPR}}
\put(-2,12.5){\rotatebox{90}{\makebox(0,0)[b]{$\overline{|\mathbf{u}|}$}}}
\end{overpic}\hfill
\begin{overpic}[width=0.46\linewidth]{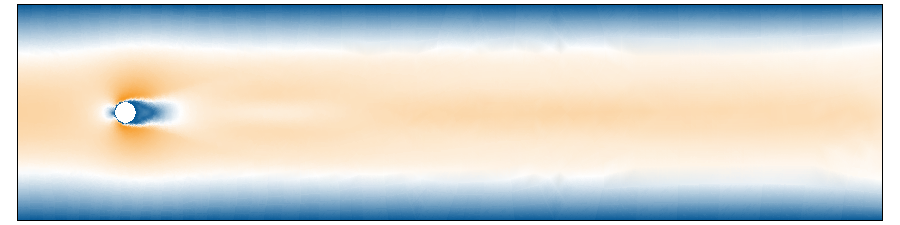}\small
\put(50,26.5){\makebox(0,0)[b]{POD-GPR}}
\end{overpic}\\[-2pt]
\includegraphics[width=0.55\linewidth]{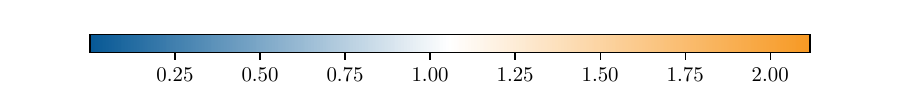}\\[4pt]
\hspace*{1.3em}\begin{overpic}[width=0.46\linewidth]{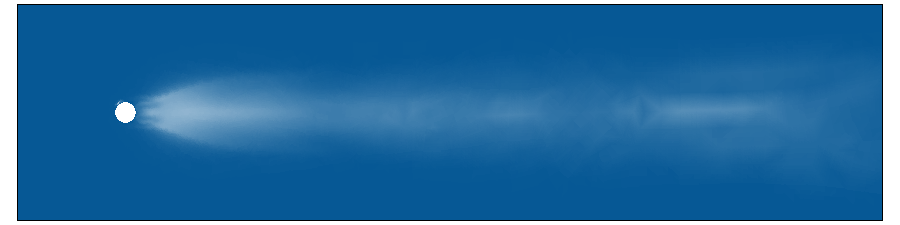}\small
\put(-2,12.5){\rotatebox{90}{\makebox(0,0)[b]{NMAE}}}
\end{overpic}\hfill
\includegraphics[width=0.46\linewidth]{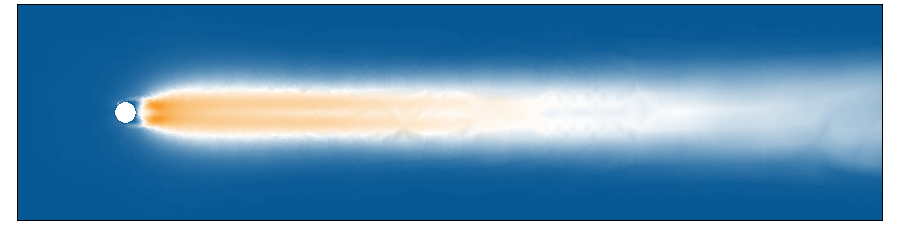}\\[-2pt]
\includegraphics[width=0.55\linewidth]{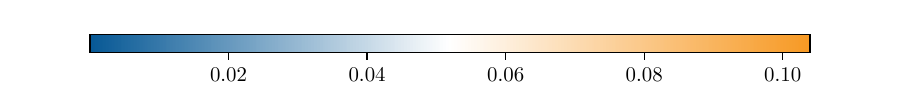}\\[4pt]
\hspace*{1.3em}\begin{overpic}[width=0.46\linewidth]{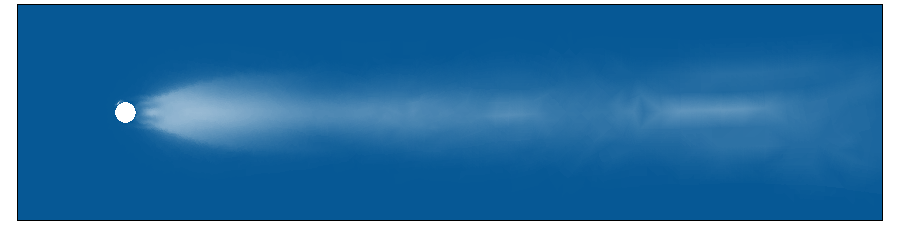}\small
\put(-2,12.5){\rotatebox{90}{\makebox(0,0)[b]{NRMSE}}}
\end{overpic}\hfill
\includegraphics[width=0.46\linewidth]{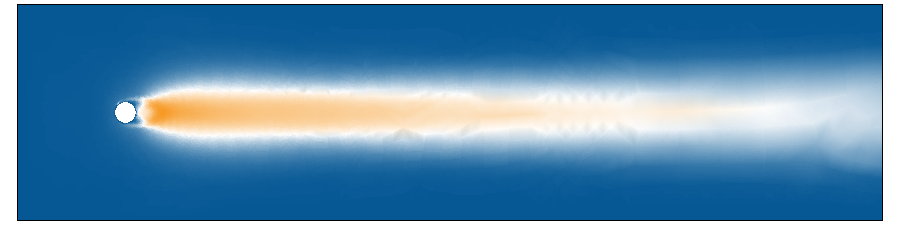}\\[-2pt]
\includegraphics[width=0.55\linewidth]{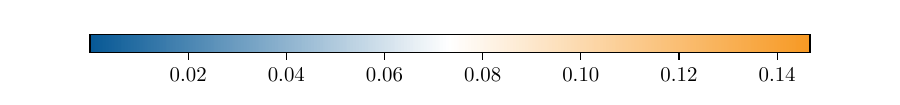}\\[4pt]
\hspace*{1.3em}\begin{overpic}[width=0.46\linewidth]{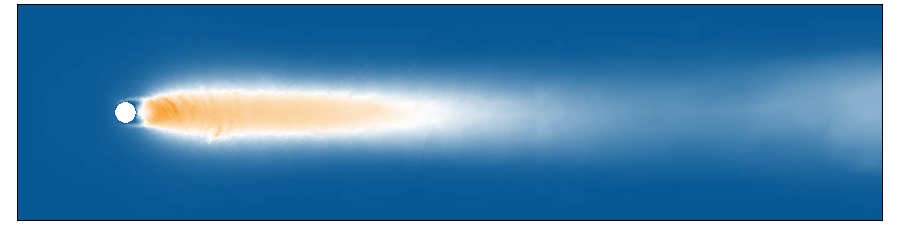}\small
\put(-2,12.5){\rotatebox{90}{\makebox(0,0)[b]{NSD}}}
\end{overpic}\hfill
\includegraphics[width=0.46\linewidth]{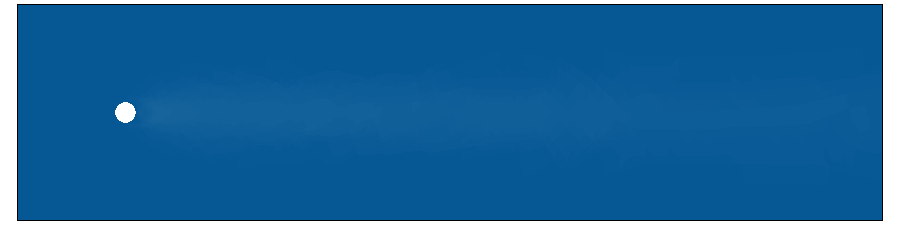}\\[-2pt]
\includegraphics[width=0.55\linewidth]{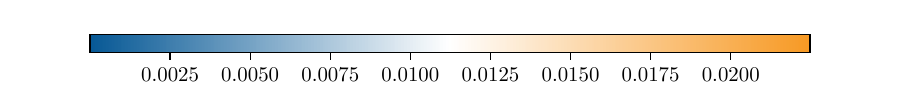}\\[4pt]
\hspace*{1.3em}\begin{overpic}[width=0.46\linewidth]{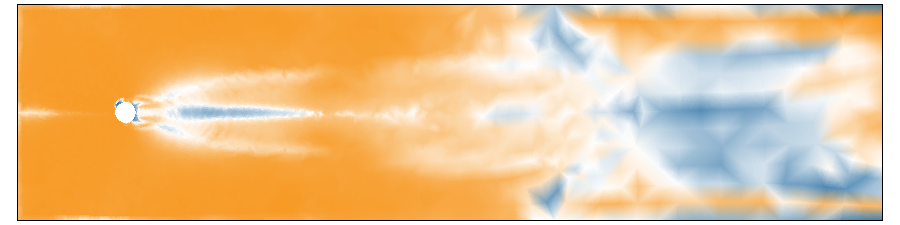}\small
\put(-2,12.5){\rotatebox{90}{\makebox(0,0)[b]{COV}}}
\end{overpic}\hfill
\includegraphics[width=0.46\linewidth]{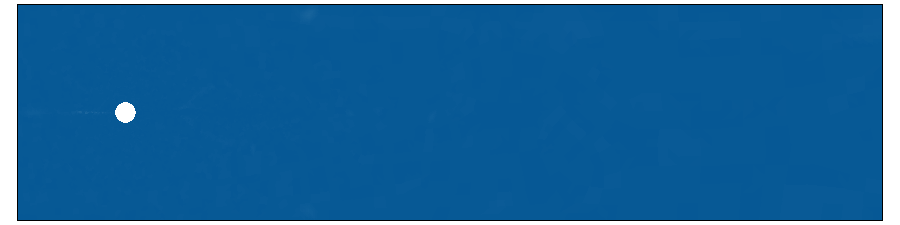}\\[-2pt]
\includegraphics[width=0.55\linewidth]{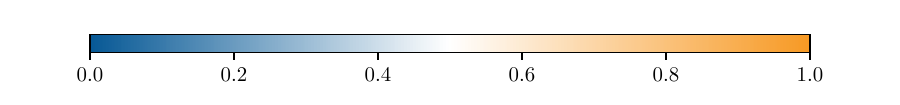}
\caption{Flow past a cylinder, Reynolds-number parametrisation: time-averaged maps covering the test window and the test parameters{; NSD denotes the normalised predicted standard deviation and COV the empirical $95\%$ coverage}. The colour scale shown below each row is shared by the maps in that row (and, for the first row, by the top map).}
\label{fig:cyl-maps}
\end{figure}

The results are given in \Cref{fig:cyl-time,fig:cyl-maps}. Both surrogate models reproduce the mean flow, but SPOD-GPR is markedly more accurate: its value of $\langle\overline{|u|}\rangle$ is within $0.2\%$ of that of the reference, compared with $2.6\%$ for POD-GPR, and it reduces the NMAE and NRMSE by about $76\%$ and $72\%$ respectively. In both cases, the errors are mostly found in the near wake, the region where vortex shedding takes place and where sensitivity to the parameters is greatest. The uncertainty estimates also differ sharply: the prediction interval of POD-GPR is far too narrow (the half-width is $6.6\times10^{-4}$ as compared with an NRMSE of $5.9\times10^{-2}${, i.e.\ NSD/NRMSE${}\approx6\times10^{-3}$}) and includes only $0.6\%$ of the test cases, which shows that it is severely overconfident. On the other hand, the $95\%$ half-width of SPOD-GPR is comparable to its NRMSE ($1.6\times10^{-2}$ in both cases), i.e.\ its predicted standard deviation is about half the actual error {(NSD/NRMSE${}\approx0.5$)}, and its interval reaches $72\%$ coverage: below the nominal $95\%$, but far better calibrated than POD-GPR. Its predicted standard deviation is concentrated in the near wake, exactly where the errors are located (\Cref{fig:cyl-maps}), so it serves as a practical indicator of error. {It should be noted, however, that the space-averaged coverage also includes the regions outside the wake, where both the error and the predicted variance are small and the coverage is close to one; within the wake, and in particular in its downstream part, the coverage is lower (\Cref{fig:cyl-maps}, bottom row).}

\subsection{Flow past an elliptic obstacle: geometric parametrisation}
\label{sec:ffd}

\begin{figure}[!htbp]
    \centering
    \includegraphics[width=0.85\linewidth]{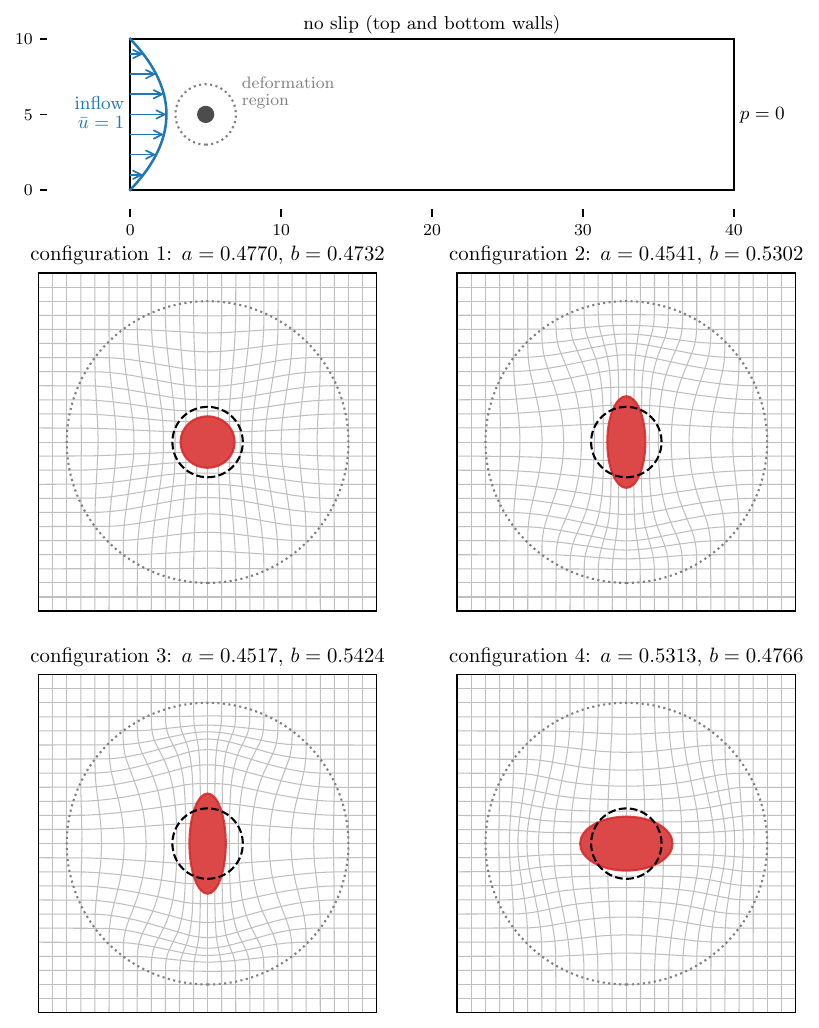}
    \caption{Flow past an elliptic obstacle. Top: the computational domain together with the boundary conditions, the reference cylinder, and the deformation region ($\rho\le 4r_0$, dotted). Bottom: four of the test configurations, including the elliptic obstacle (red), the reference circle (dashed), and a Cartesian grid pushed forward by the deformation map. To improve visibility, the deviation of the ellipse from the circle, $(a-r_0,\,b-r_0)$, is magnified five times in the drawings; the actual semi-axes are listed above each panel.}
    \label{fig:ellipse-configs}
\end{figure}

\begin{figure}[tbp]
\centering
\vspace*{1.2em}
\begin{overpic}[width=0.5\textwidth]{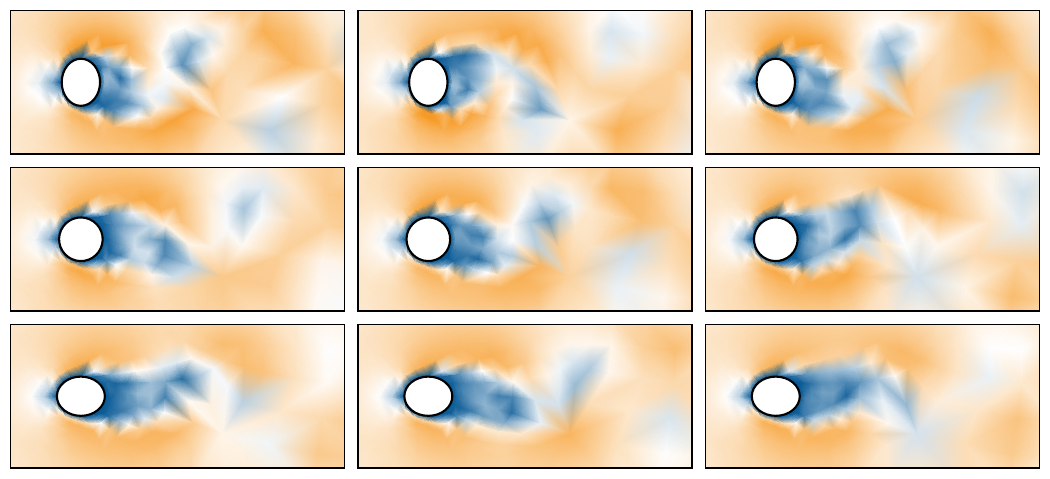}
\small
\put(16.8,46.5){\makebox(0,0)[b]{$t=0$}}
\put(49.85,46.5){\makebox(0,0)[b]{$t=50$}}
\put(82.95,46.5){\makebox(0,0)[b]{$t=100$}}
\footnotesize
\put(-1,37.6){\makebox(0,0)[r]{\shortstack[r]{$a=0.45$\\$b=0.55$}}}
\put(-1,22.6){\makebox(0,0)[r]{\shortstack[r]{$a=0.50$\\$b=0.50$}}}
\put(-1,7.75){\makebox(0,0)[r]{\shortstack[r]{$a=0.55$\\$b=0.45$}}}
\end{overpic}

\vspace{22pt}
\begin{overpic}[width=0.82\textwidth]{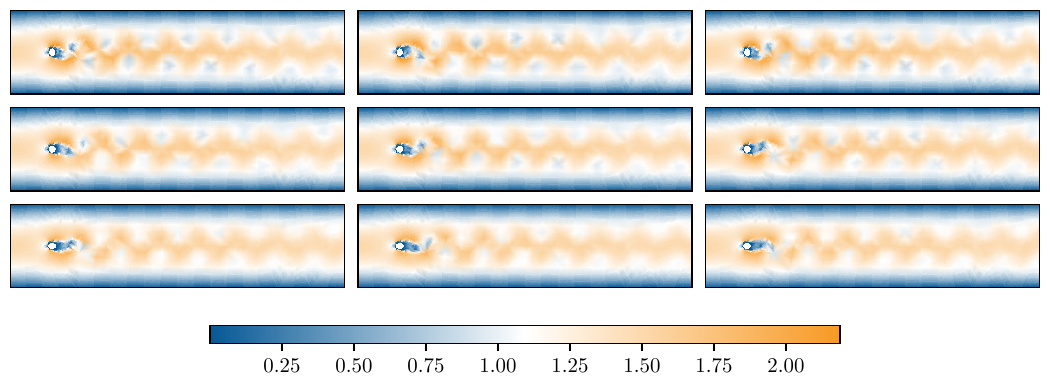}
\small
\put(16.8,37){\makebox(0,0)[b]{$t=0$}}
\put(49.85,37){\makebox(0,0)[b]{$t=50$}}
\put(82.95,37){\makebox(0,0)[b]{$t=100$}}
\footnotesize
\put(-0.5,31.15){\makebox(0,0)[r]{\shortstack[r]{$a=0.45$\\$b=0.55$}}}
\put(-0.5,21.95){\makebox(0,0)[r]{\shortstack[r]{$a=0.50$\\$b=0.50$}}}
\put(-0.5,12.75){\makebox(0,0)[r]{\shortstack[r]{$a=0.55$\\$b=0.45$}}}
\small
\put(81,4.25){\makebox(0,0)[l]{$|\mathbf{u}|$}}
\end{overpic}
\caption{Flow past an elliptic obstacle, geometric parametrisation{: full-order velocity magnitude $|\mathbf{u}|$ on the reference configuration for three values of $(a,b)$ (rows) at three time instants (columns), with time measured from the first stored snapshot; top: close-up of the near-wake region, bottom: full domain}.}
\label{fig:ffd-maps-time}
\end{figure}

The third test case uses the same channel shape and numerical solver at a fixed $Re=100$, but substitutes the circular cylinder of radius $r_0=0.5$ with an ellipse which has semi-axes $(a,b)$. A two-parameter vector $\boldsymbol{\mu}=(a,b)\in[0.45,0.55]^2$ is thus defined. The values of $(a,b)$ are selected uniformly from this square; altogether $100$ different geometries are generated, of which $80$ are used for training and $20$ are kept for testing. The change in geometry is applied to both the computational mesh and the output mesh by means of a smooth radial transformation,
\begin{equation}
\mathbf{x}\mapsto \mathbf{c}+\operatorname{diag}\!\big(\beta(\rho)\tfrac{a}{r_0}+1-\beta(\rho),\ \beta(\rho)\tfrac{b}{r_0}+1-\beta(\rho)\big)(\mathbf{x}-\mathbf{c}),\qquad \rho=|\mathbf{x}-\mathbf{c}|,
\end{equation}
where $\mathbf{c}$ represents the centre of the obstacle and $\beta$ is a $C^2$ quintic smoothstep function. The function $\beta$ has a value of $1$ at the boundary of the obstacle ($\rho\le r_0$), so that the circle is mapped precisely onto the ellipse, and reaches $0$ when $\rho\ge 4r_0$, where the mapping reduces to the identity. As the mesh connectivity is preserved by this mapping, the velocity field calculated in each deformed domain is stored at the nodes of the \emph{reference} mesh, which ensures that all the snapshots have the same degrees of freedom and possess the same weight matrix (that is, the lumped mass matrix of the reference mesh). This is in line with the standard approach for geometrically parametrised ROMs, wherein the solution manifold is represented on a fixed reference configuration. The channel set-up and four of the test geometries are shown in \Cref{fig:ellipse-configs}{, and representative snapshots of the velocity magnitude in \Cref{fig:ffd-maps-time}}.

\begin{figure}[t]
    \centering
    \includegraphics[width=\linewidth]{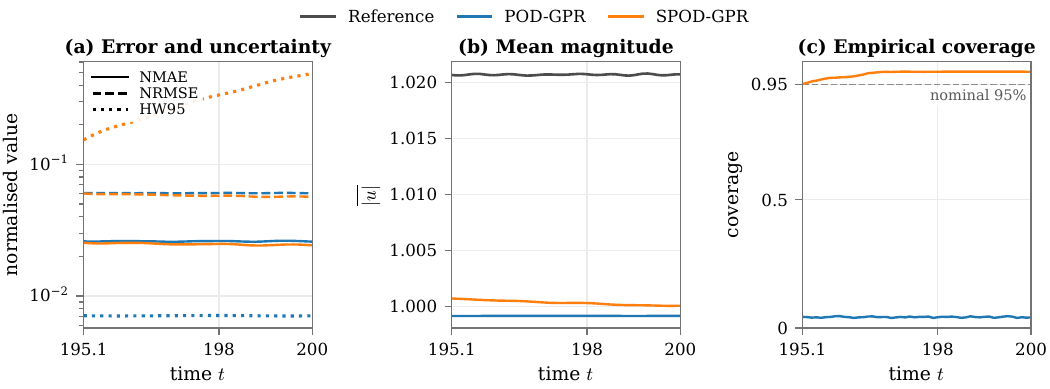}
    \caption{Flow past an elliptic obstacle, geometric parametrisation, test window. {On the left: the NMAE (solid), the NRMSE (dashed) and the predicted normalised $95\%$ half-width (dotted) for SPOD-GPR (orange) and POD-GPR (blue); for a calibrated Gaussian predictive distribution, the half-width is about $1.96$ times the NRMSE. In the centre: the space average of the mean velocity magnitude $\overline{|\mathbf{u}|}$, with the reference solution in black. On the right: the empirical coverage of the $95\%$ predictive interval.}}
    \label{fig:ffd-time}
\end{figure}

\begin{figure}[p]
\centering
\vspace*{1.2em}
\begin{overpic}[width=0.46\linewidth]{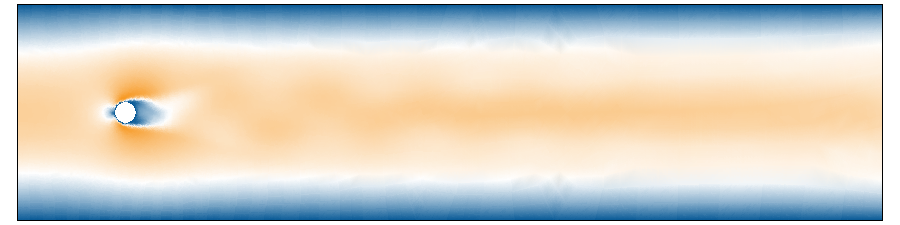}\small
\put(50,26.5){\makebox(0,0)[b]{Reference}}
\put(-2,12.5){\rotatebox{90}{\makebox(0,0)[b]{$\overline{|\mathbf{u}|}$}}}
\end{overpic}\\[14pt]
\hspace*{1.3em}\begin{overpic}[width=0.46\linewidth]{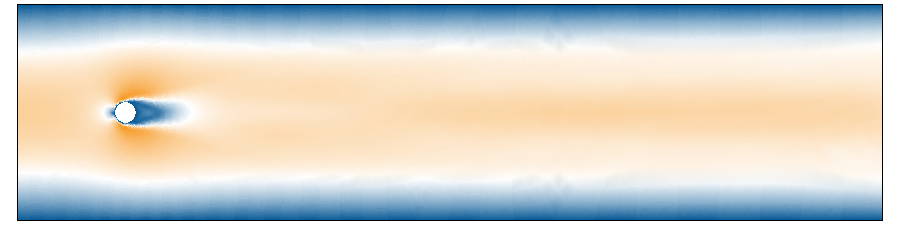}\small
\put(50,26.5){\makebox(0,0)[b]{SPOD-GPR}}
\put(-2,12.5){\rotatebox{90}{\makebox(0,0)[b]{$\overline{|\mathbf{u}|}$}}}
\end{overpic}\hfill
\begin{overpic}[width=0.46\linewidth]{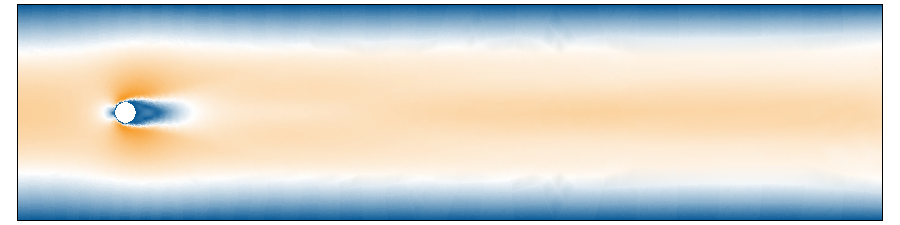}\small
\put(50,26.5){\makebox(0,0)[b]{POD-GPR}}
\end{overpic}\\[-2pt]
\includegraphics[width=0.55\linewidth]{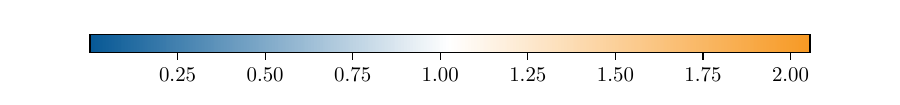}\\[4pt]
\hspace*{1.3em}\begin{overpic}[width=0.46\linewidth]{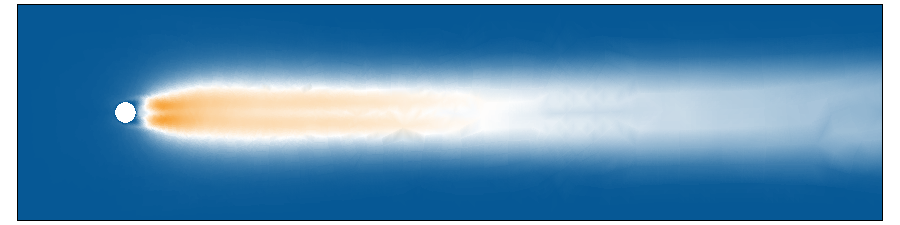}\small
\put(-2,12.5){\rotatebox{90}{\makebox(0,0)[b]{NMAE}}}
\end{overpic}\hfill
\includegraphics[width=0.46\linewidth]{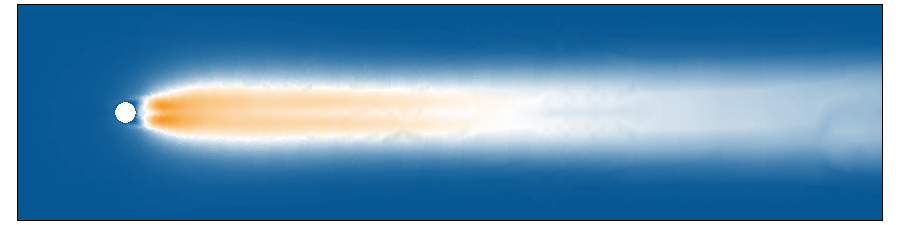}\\[-2pt]
\includegraphics[width=0.55\linewidth]{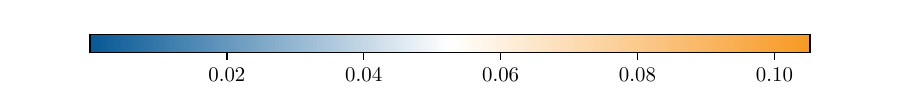}\\[4pt]
\hspace*{1.3em}\begin{overpic}[width=0.46\linewidth]{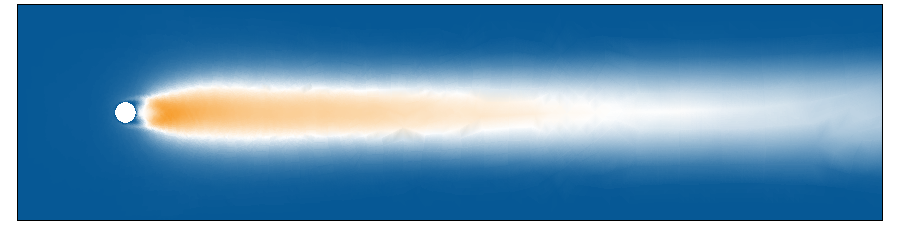}\small
\put(-2,12.5){\rotatebox{90}{\makebox(0,0)[b]{NRMSE}}}
\end{overpic}\hfill
\includegraphics[width=0.46\linewidth]{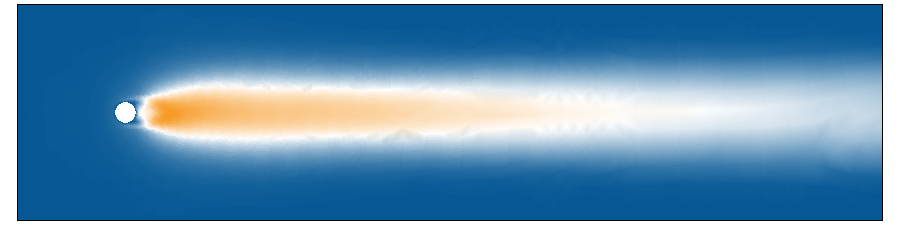}\\[-2pt]
\includegraphics[width=0.55\linewidth]{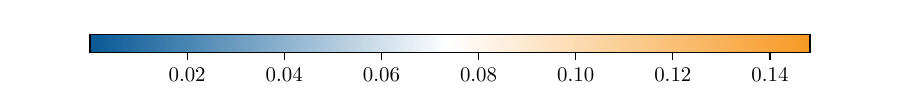}\\[4pt]
\hspace*{1.3em}\begin{overpic}[width=0.46\linewidth]{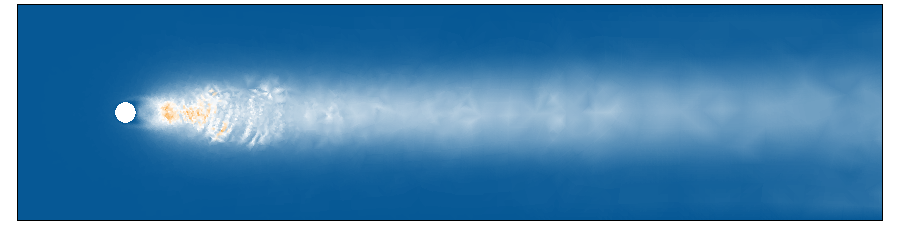}\small
\put(-2,12.5){\rotatebox{90}{\makebox(0,0)[b]{NSD}}}
\end{overpic}\hfill
\includegraphics[width=0.46\linewidth]{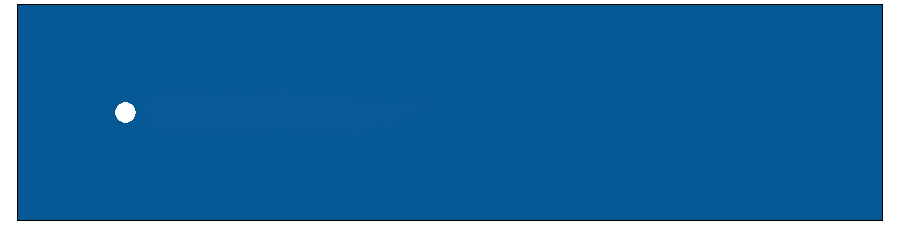}\\[-2pt]
\includegraphics[width=0.55\linewidth]{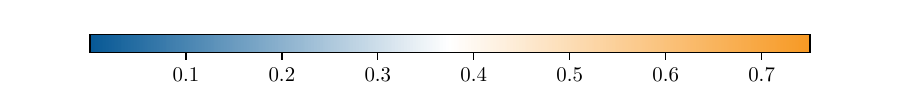}\\[4pt]
\hspace*{1.3em}\begin{overpic}[width=0.46\linewidth]{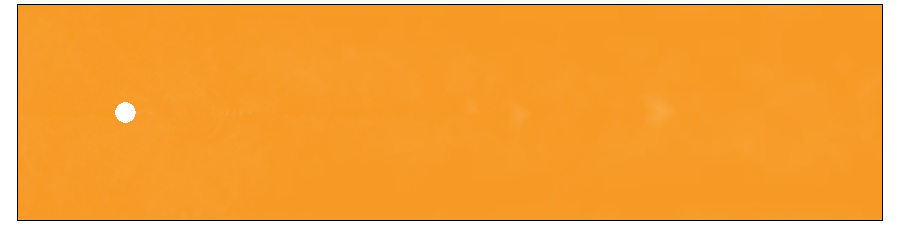}\small
\put(-2,12.5){\rotatebox{90}{\makebox(0,0)[b]{COV}}}
\end{overpic}\hfill
\includegraphics[width=0.46\linewidth]{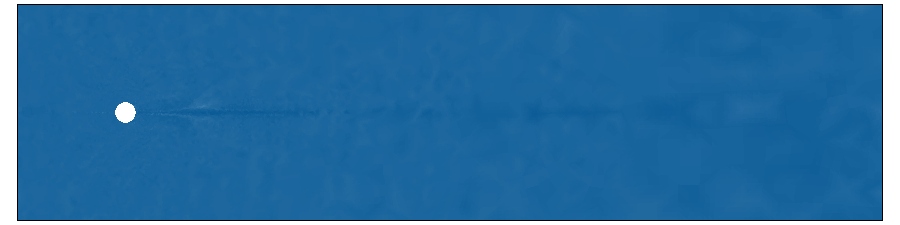}\\[-2pt]
\includegraphics[width=0.55\linewidth]{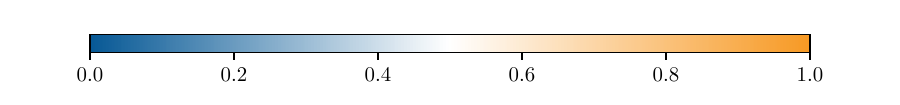}
\caption{Flow past an elliptic obstacle, geometric parametrisation: the time-averaged maps over the test window and test parameters, presented on the reference configuration{; NSD denotes the normalised predicted standard deviation and COV the empirical $95\%$ coverage}. The colour scale shown below each row is shared by the maps in that row (and, for the first row, by the top map).}
\label{fig:ffd-maps}
\end{figure}

The results are presented in \Cref{fig:ffd-time,fig:ffd-maps} and \Cref{tab:results}. This set-up is more difficult than the previous one since the geometry changes the near-wake topology and the shape of the obstacle influences both the shedding frequency and phase; the errors {of SPOD-GPR} are about four times greater than in the Reynolds case {(those of POD-GPR are comparable)} and are mostly concentrated in the near wake located downstream of the obstacle. Both models slightly underestimate the velocity magnitude (by around $2\%$). SPOD-GPR reduces the NMAE and NRMSE by approximately $5\%$ and $4\%$ respectively. As in the Reynolds case, the POD-GPR uncertainty is highly overconfident (with a coverage of $4\%${, NSD/NRMSE${}\approx0.06$}), while the SPOD-GPR interval widens as the forecast horizon increases, as is expected for an autoregressive extrapolation, and is conservative (half-width $0.33$ against an NRMSE of $0.058${, i.e.\ NSD/NRMSE${}\approx2.9$}), with an average coverage of $0.99${; by the end of the test window, however, the half-width approaches half of the data range, so that the band becomes only weakly informative}.

\FloatBarrier
\section{Discussion}
\label{sec:discussion}

\paragraph{Accuracy.} On the four benchmark problems, SPOD-GPR reduces the NMAE by $5$--$76\%$ and the NRMSE by $4$--$72\%$ compared to POD-GPR. The gains are largest for the Reynolds-parametrised cylinder flow and smallest for the elliptic obstacle. In the periodic Kuramoto--Sivashinsky scenario, the benefit is mainly qualitative: the time-domain autoregressive model of POD-GPR suppresses the oscillations after just a few steps, while the spectral formulation—where a periodic waveform is represented by slowly varying coefficients at a small number of fixed frequencies—keeps the oscillations going for the whole prediction period. This is in agreement with the mechanism described in \Cref{sec:motivation}: a single oscillatory scalar coefficient is not autonomous, but its windowed Fourier coefficients are. It should be noted that, with the set-up in \Cref{tab:hyper}, the dynamics GP in SPOD-GPR has a latent dimension of $2rK$ ($6$ and $160$ in the two KS cases and $12$ for the flows) as opposed to $r_{\mathrm{POD}}$ for POD-GPR ($2$, $6$, and $2$), so the spectral approach also makes use of a higher-dimensional latent state. Nevertheless, the comparison is still valid since the latent dimension of POD-GPR cannot be increased without limit: the POD coefficients at higher indices become more irregular in time and hence harder to predict.

\paragraph{Uncertainty.} The key difference between the two methods is the way in which they quantify uncertainty. In both flow cases, the prediction bands from POD-GPR are between about one and two orders of magnitude narrower than the actual error and cover only $0.6$--$4\%$ of the test samples. A possible explanation is as follows. {In both methods, the variance of the dynamics GP is calibrated on the leave-one-parameter-out residuals of single-step increments}; in the time domain, the changes between consecutive snapshots are very small and therefore easy to predict, so the calibration fails to account for the accumulation of phase error over a long rollout. On the other hand, in the spectral case each autoregressive step moves the entire windowed block to the next position; the coefficients change more slowly and smoothly, and {the same} calibration more accurately represents the true extrapolation error. Moreover, the uncertainty associated with the SPOD modes—those that contain phase information—is passed on to the reconstructed field. As a result, SPOD-GPR produces intervals that range from moderately overconfident (cylinder, $72\%$) to conservative (elliptic obstacle, $99\%$){, with spread--skill ratios between $0.5$ and $2.9$ (against $6\times10^{-3}$ and $6\times10^{-2}$ for POD-GPR),} and exhibit a spatial pattern that matches the error. With regard to the KS equation, SPOD-GPR is only slightly overconfident in the travelling-wave regime ($88\%$ coverage, with intervals about half the width of those obtained from POD-GPR) and more overconfident in the periodic regime ($77\%$), where POD-GPR is instead conservative ($99\%$) but with intervals about five times wider than {those of SPOD-GPR}. Therefore, improving the calibration of the dynamics GP for long, near-periodic rollouts should be a major focus of future work.

\paragraph{Limitations.} The test windows employed in this study are short compared with the training window ($5\%$ of the snapshots), and the ranges of parameters investigated are only moderately wide. For longer prediction horizons, the autoregressive rollout will eventually drift in phase and the posterior mean will tend to revert to the mean flow, as is already indicated by the amplitude underestimation observed in all of the KS tests. The method also relies on a number of hyperparameters (the POD rank $R$, the window length $N_f$, the number of frequencies $K$, the SPOD rank $r$, the kernel length scales), which have been selected manually in this case. Although using a sliding window with $h=1$ gives the highest number of training pairs for the dynamics GP, it also results in a considerable amount of redundancy and thus increases the computational cost of the Nyström regression. Finally, the separate SVDs and Procrustes alignments for each parameter assume that the dominant subspaces change smoothly as the parameter varies, an assumption which can fail near or across bifurcations.

\section{Conclusions}
\label{sec:conclusions}

We have developed SPOD-GPR, a non-intrusive, parametric, probabilistic spatio-temporal reduced-order model which makes use of the frequency structure of the data{: the regression is performed on frequency-resolved SPOD coefficients rather than on time-domain POD coefficients}. SPOD is motivated by augmenting the state with time delays: applying a Fourier transform along the delay axis separates the temporal scales that would otherwise be mixed up by block POD, and as a result temporally coherent modes are obtained, the coefficients of which are easier to learn. A first step of POD reduction is carried out in order to make the subsequent spectral analysis computationally feasible. The parameter dependence of all the bases is then learned on a Grassmann manifold chart following a Procrustes alignment, thus making sure that the predicted modes stay exactly orthonormal. The modal coefficients are obtained by means of a Gaussian process that maps parameters to trajectories and a parameter-conditioned autoregressive Gaussian process, and the uncertainty is passed on to the physical field through pathwise sampling of all the GPs.

{The method was tested on} four benchmark problems{:} the two-dimensional Kuramoto--Sivashinsky equation in both periodic and travelling-wave regimes, flow past a cylinder parametrised by the Reynolds number, and flow past an elliptic obstacle parametrised by its semi-axes. {In all of them, for} unseen parameters and future times, {it} improved on the accuracy of a POD-GPR model using the same regression architecture. Most importantly, for the flow cases it also produced uncertainty estimates {that were far better calibrated than those of} POD-GPR, which was strongly overconfident. For the periodic Kuramoto--Sivashinsky case, SPOD-GPR kept the temporal oscillations that POD-GPR failed to capture, although its intervals were overconfident, while those of POD-GPR were conservative but much wider. {In the travelling-wave regime, instead, the predicted standard deviation of SPOD-GPR closely matched its actual error, whereas POD-GPR was conservative.} Future work will concentrate on the automated tuning of the spectral hyperparameters, extending the forecast horizons, {improving the calibration of the dynamics GP for long, near-periodic rollouts,} dealing with higher-dimensional parameter spaces, and making use of the calibrated uncertainty in uncertainty-aware ROMs and in real-time digital-twin applications.

\bibliographystyle{unsrtnat}
\bibliography{refs}

\end{document}